\documentclass[lettersize,journal]{IEEEtran}

\usepackage{subcaption}

\usepackage{hyperref}
\usepackage{amsmath}

\usepackage{amssymb}

\usepackage{lineno}

\usepackage{booktabs,caption}
\usepackage[flushleft]{threeparttable}
\usepackage{amsfonts}
\usepackage{float}
\usepackage{placeins}
\usepackage{textcomp}
\usepackage{amsmath}
\usepackage{graphicx}
\usepackage{url}
\usepackage{multicol, blindtext}
\usepackage{booktabs}
\usepackage[table]{xcolor}
\usepackage{xcolor}
\usepackage{algorithmicx}
\usepackage{lipsum}

\usepackage{amssymb}
\usepackage{lineno}
\usepackage{algpseudocode}
\usepackage{amsfonts}
\usepackage{siunitx}
\usepackage{amsmath}
\usepackage{multirow}
\usepackage{wrapfig}
\usepackage{array}
\usepackage{color}
\usepackage{silence}
\usepackage[table]{xcolor}
\definecolor{lightgray}{gray}{0.9}
\usepackage[linesnumbered,ruled,vlined]{algorithm2e}
\usepackage{tabularx}
\usepackage{amssymb}
\usepackage{lineno}

\usepackage[english]{babel}
\usepackage{ragged2e}
\usepackage{blindtext}
\usepackage{booktabs,caption}
\usepackage[flushleft]{threeparttable}
\usepackage{amsfonts}
\usepackage{float}
\usepackage{placeins}
\usepackage{textcomp}
\usepackage{amsmath}
\usepackage{graphicx}
\usepackage{url}
\usepackage{multicol, blindtext}
\usepackage{booktabs}
\usepackage[table]{xcolor}
\usepackage{xcolor}
\usepackage{lipsum}

\usepackage{amssymb}
\usepackage{lineno}
\usepackage{algpseudocode}
\usepackage{amsfonts}
\usepackage{siunitx}
\usepackage{amsmath}
\usepackage{multirow}
\usepackage{wrapfig}
\usepackage{array}
\usepackage{color}
\usepackage{silence}
\usepackage[table]{xcolor}
\definecolor{lightgray}{gray}{0.9}
\usepackage{svg}

\begin{document}

\title{CIC-Trap4Phish: A Unified Multi-Format Dataset for Phishing and Quishing Attachment Detection}

\author{Fatemeh Nejati, Mahdi Rabbani, Morteza Eskandarian,  Mansur Mirani, Gunjan Piya, Igor Opushnyev, Ali A. Ghorbani, ~\IEEEmembership{Senior Member, IEEE}, Sajjad Dadkhah,~\IEEEmembership{Senior Member, IEEE}
\thanks{F. Nejati, M. Rabbani, M Eskandarian, A A. Ghorbani, and S. Dadkhah are with the Canadian Institute for Cybersecurity, Faculty of Computer Science, University of New Brunswick, Fredericton, NB, Canada (e-mails: \{fatemeh.nejati, m.rabbani, morteza.eskandarian, ghorbani, sdadkhah\}@unb.ca). M. Mirani, G. Piya, I. Opushnyev are with the Mastercard Vancouver Tech Hub, Vancouver, British Columbia, Canada (e-mails: \{mansur.mirani, sunny.piya, igor.opushnyev\}@mastercard.com).}}

\markboth{}
{Shell \MakeLowercase{\textit{et al.}}: A Sample Article Using IEEEtran.cls for IEEE Journals}


\maketitle

\begin{abstract}
Phishing attacks represents one of the primary attack methods which is used by cyber attackers. In many cases, attackers use deceptive emails along with malicious attachments to trick users into giving away sensitive information or installing malware while compromising entire systems. The flexibility of malicious email attachments makes them stand out as a preferred vector for attackers as they can embed harmful content such as malware or malicious URLs inside standard document formats. Although phishing email defenses have improved a lot, attackers continue to abuse attachments, enabling malicious content to bypass security measures. Moreover, another challenge that researches face in training advance models, is lack of an unified and comprehensive dataset that covers the most prevalent data types. To address this gap, we generated CIC-Trap4Phish, a multi-format dataset containing both malicious and benign samples across five categories commonly used in phishing campaigns: Microsoft Word documents, Excel spreadsheets, PDF files, HTML pages, and QR code images. For the first four file types, a set of execution-free static feature pipeline was proposed, designed to capture structural, lexical, and metadata-based indicators without the need to open or execute files. Feature selection was performed using a combination of SHAP analysis and feature importance, yielding compact, discriminative feature subsets for each file type. The selected features were evaluated by using lightweight machine learning models, including Random Forest, XGBoost, and Decision Tree. All models demonstrate high detection accuracy across formats. For QR code-based phishing (quishing), two complementary methods were implemented: image-based detection by employing Convolutional Neural Networks (CNNs) and lexical analysis of decoded URLs using recent lightweight language models.

\end{abstract}

\begin{IEEEkeywords}
Malicious Attachments, QR Code Phishing, Static Feature Extraction, Lightweight Language Models, Email Security, Benchmark Dataset
\end{IEEEkeywords}

\section{Introduction}
Organizations use different file formats, such as PDF, Word, Excel, and HTML, as primary tools for data exchange and communication across their operations \cite{liu2025analyzing}. In parallel, Quick Response (QR) codes have become a widely used mechanism for seamless website access, which enable users to establish web connections through encoded URLs without manual input \cite{alsuhibany2025innovative}.Therefore, these file formats represent the most common types of email attachments.

\begin{figure}[t]
  \centering
  \includegraphics[width=0.8\columnwidth]{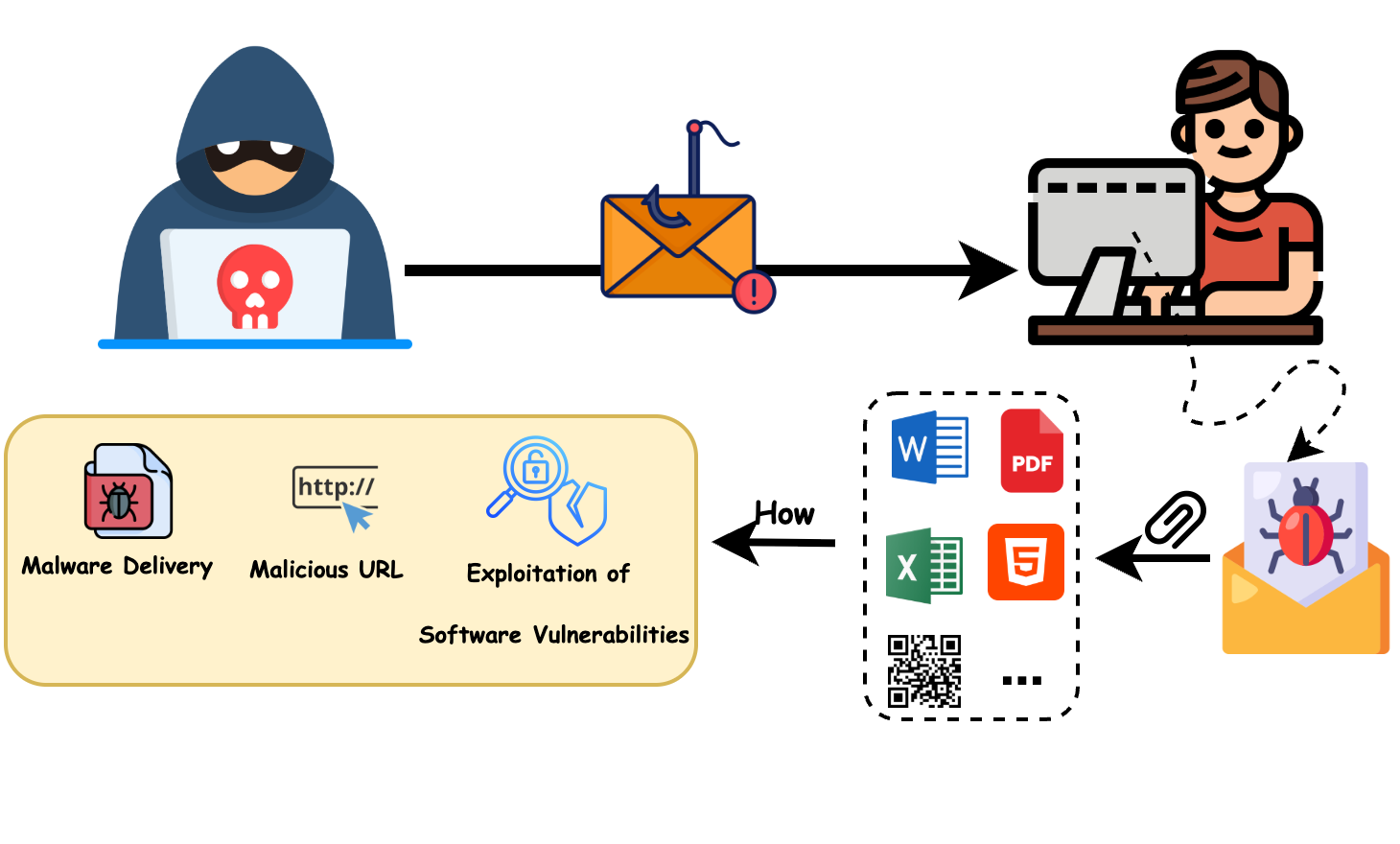}
  \caption{Sources of collected phishing URLs.}
  \label{fig:phishingSenario}
\end{figure}

However, their popular use and flexible structure have made them an attractive tool for distributing malicious content, enabling attackers to hide scripts, macros, hyperlinks, embedded objects, and malicious metadata within email attachments that appear legitimate \cite{liu2025analyzing}. Such hidden threats can be activated when a user opens a file, clicks an embedded link, or scans a QR code (quishing) \cite{trad2025detecting} that allows the attacker to deliver the malicious payload.
Beyond traditional \textit{quishing} attacks that lead victims to malicious websites, there is an advance kind of \textit{quishing}, called \textit{QRLJacking} (Quick Response Code Login Jacking). In this kind of attack, in order to steal victims' login sessions  and access users' accounts without using any passwords, attackers use a legitimate QR-based login system \cite{zhang2025demystifying} and trick users to scan the fake QR codes. These kinds of attacks prove the urgent need for providing strong security measures.
Therefore, detecting threats embedded in diverse file types requires strong mechanisms to enhance overall cybersecurity resilience.

Detecting these types of attacks is technically challenging because email content and sender addresses often mimic legitimate communications encountered by end users. Recent advances in large language models (LLMs) allow attackers to generate highly convincing, contextually coherent messages that can evade traditional keyword-based phishing detectors \cite{heiding2024evaluating}, bringing attackers one step closer to successful compromise. Figure \ref{fig:phishingSenario} demonstrates how attackers use malicious attachments to carry out their attacks in a typical phishing sequence. Another major challenge in detecting and analyzing such malware threats lies in the type of analysis required to uncover their hidden behaviors. Dynamic analysis typically provides deeper visibility into malicious attachments because it reveals internal structure, runtime function calls, API usage, and control-flow behavior \cite{wagener2008malware}. However, it requires isolated execution environments, significant computational resources, and careful containment mechanisms to avoid live payload execution, which makes the deployment computationally expensive and operationally impractical inside email gateways. For these reasons, static analysis that extracts safe, non-executing features from attachment files provides more efficient and safer alternative which avoids execution risk and latency while enabling high-throughput screening of incoming mail \cite{noman2021static}.

Apart from detection methodologies, the malicious attachments domain faces additional challenges due to the lack of comprehensive, high-quality datasets that cover the most prevalent file types. Existing datasets are often limited to specific file types, fragmented, and fail to include he broad range of file types commonly found in email attachments \cite{lu2021universal}. This limitation restricts the development and evaluation of robust detection models that are capable of generalizing across heterogeneous attachment file formats encountered in real-world phishing campaigns.

To address this critical gap, we propose CIC-Trap4Phish, an unified multi-format dataset that contains both malicious and benign samples across five file formats: Word documents, Excel spreadsheets, PDF files, HTML pages, and QR codes.  Malicious samples were collected from reputable sources including Malware Bazaar \cite{malwarebazaar2024}, PhishTank \cite{phishtank2024}, PDFMal2022 dataset \cite{issakhani2022pdf},  Nazario Phishing Email Corpus \cite{gonzalez2011phishing}, Chakraborty Phishing Dataset \cite{subhajournal_phishingemails}, and other trusted repositories to ensure coverage of all targeted file types. To collect benign samples, we used multiple strategies including generating Excel spreadsheets and QR code samples, crawling Word documents and HTML pages from trusted sources such as Google and Wikipedia, and also using samples from available datasets. The generated dataset provides researchers with a balanced and diverse benchmark for studying attachment-based threats across multiple file types.

For extracting and selecting features across the generated samples, several customized approaches were designed based on the characteristics of each document type. For all four categories, Word documents, Excel spreadsheets, PDF files, and HTML pages, a diverse set of structural, static, and content-based attributes were extracted. To further reduce feature dimensionality, feature importance techniques integrated with explainability-based analysis were applied to identify the most influential and relevant attributes within each document category, resulting in the selection of the top 10 features per type. The selected features were then evaluated using lightweight machine learning classifiers, and the results showed strong performance in distinguishing malicious and benign samples. For QR code analysis, two statistical approaches were used. First, a Convolutional Neural Network (CNN) method was applied to QR code images to extract spatial and pattern-based features. Second, the embedded URL strings were decoded from the QR codes, and lexical analysis was performed on the extracted words and characters. These tokens were then processed through lightweight language models such as BERT-Tiny, DeBERTa-v3, ModernBERT, and DeepSeek-R1 to identify the most efficient model with low computational cost and high detection accuracy.

To the best of our knowledge, this research presents the first comprehensive and unified phishing attachment dataset that encompasses all prevalent file types. In addition to collecting and generating file samples, the study introduces a set of static features extracted directly from the raw files across the four previously mentioned data types. The main contributions of this paper are summarized as follows:
\begin{itemize}
    \item We generated a unified and complete repository which contains both benign and malicious files that are commonly leveraged for delivering phishing content. The dataset includes multiple file types such as Word documents, Excel spreadsheets, PDFs, HTML files, and QR codes.  
    \item We proposed efficient and lightweight static features for each file category, eliminating the need to open or execute files. This design can significantly enhance analysis speed and safety in real-world security applications.
    \item We conducted a comprehensive evaluation to assess the quality of the extracted features using lightweight machine learning approaches. Additionally, for QR code detection, two complementary methods were proposed: one based on image-based detection, and the other on lexical analysis of the embedded URL links using recent lightweight language models, including BERT-Tiny, DeBERTa-v3, ModernBERT, and Deep Seek-R1.
\end{itemize}

The remainder of this paper is organized as follows. Section \ref{relatedwork_Nazgol_new} reviews existing works and related datasets. Section \ref{Nazgol_Malware_methodology_Nazgol_new} describes the steps involved in generating the proposed dataset, including data collection. Section 4 explains the processes of feature extraction, feature selection, and classification. Section \ref{Nazgol_Malware/Evaluation_new} presents the experimental results and evaluation criteria. Finally, Section \ref{Nazgol_Malware_conclusion} concludes the paper and outlines future research directions.

\section{{Related Work}}
\label{relatedwork_Nazgol_new} 
In recent years, researchers have attempted to develop preventive mechanisms against the growing misuse of email attachments by constructing specialized datasets for different file types and designing corresponding detection techniques. Several studies have focused primarily on PDF files, while the importance of other file types, particularly QR codes, which have become a major tool for quishing attacks, has been largely overlooked in the literature. This section presents recent detection techniques for malicious attachment files, reviews the available datasets, and discusses their limitations and characteristics.

Issakhani et al. \cite{issakhani2022pdf} introduce a PDF dataset called Evasive-PDFMal2022. The collected dataset contains 11,173 malicious, and 9,109 benign files. A total of 37 static features including 12 general features and 25 structural features were extracted from each file. To show effectiveness of extracted features, they proposed a stacking ensemble model that combines Random Forest, SVM, and other models. Their model achieved 99.89\% accuracy on Contagio and 98.69\% on the new dataset, which shows improved robustness against evasive PDF malware compared to prior detection methods.

Hossain et al. \cite{hossain2024pdf} proposed a robust method for detecting malicious PDF file,  emphasizing dataset generation and feature extraction. The dataset contains 15,958 PDF samples. The set of features is extracted from structural, metadata, and content layers using PDFiD, PDFINFO, and PDF-PARSER tools. The performance of these features is examined by Random Forest classifier. Moreover, in order to improve the model transparency, 23000 decision rules are extracted and are explained by SHAP (Sharpley Additive exPlanations). The results demonstrated superior accuracy and transparency compared with prior studies, including PDFMal-2022. In addition, 
the EMBER2024 dataset \cite{joyce2025ember2024} provides a large-scale benchmark that includes approximately 3.2 million files across six different formats Win32/ Win64/ .NET/ APK/ ELF/ PDF), that was collected over 64 weeks. The dataset includes VirusTotal rescanned files where benign files show zero detections after thirty days and malicious files receive five or more independent engine detections. Therefore, they were separated from training and testing data to evaluate system robustness. For evaluation, the LightGBM baseline achieves near-perfect results on the standard test split.

The authors Casino et al. \cite{casino2023analysis} introduce a new lightweight framework for detecting malicious Microsoft Office documents through their embedded visual elements. For this purpose, 14,531 malicious samples and 890 benign macro-enabled Office files  were collected. The experimental results show that the system achieves more than 99\% accuracy and 96\% F1-score and it also provides the ability to identify specific campaign origins.
In order to address the limitations of prior detection methods, Koutsokostas et al. \cite{koutsokostas2022invoice} managed a comprehensive overview of malicious Microsoft Office document. In this work they integrated both static and dynamic analysis techniques and they created a large, well-balanced dataset of over 18,000 Office files. In this dataset, 40 features were extracted and then evaluated by using different classifiers (Random Forest, XGBoost, MLP, and SVM). 

The DikeDataset \cite{iosifache_dikedataset_2023} is published as a GitHub repository that contains Office documents with malicious content (e.g. .doc/x/m, .xls/x/m, .ppt/x/m) and benign OLE files extracted from websites and public databases. The VirusTotal API generates malware labels through engine voting. The MIT license governs this dataset which provides researchers with a practical method to train and evaluate static detectors for both executable files and Office documents. This dataset does not contain any evaluation. In addition, 
Contagio Malware Dump \cite{parkour2013_contagio_counts} contains a large collection of mixed dataset that includes both benign and malicious samples across multiple formats (XLS/XLSX, DOC/DOCX, PPT/PPTX, PDF, RTF, ZIP, JAR, ELF, Mach-O). The maintainers of the repository provide a substantial collection of clean files which researchers use to prevent false-positive results. The public database contains 16,800 benign and 11,960 malicious files according to public reports.

\definecolor{rowgray}{gray}{0.95}

\begin{table*}[t]
\centering
\scriptsize
\caption{Summary of malicious attachment datasets by year, scale, file types.}
\rowcolors{2}{rowgray}{white}      
\renewcommand{\arraystretch}{1.5} 
\begin{tabular}{
        >{\centering\arraybackslash}p{0.26\linewidth}
        >{\centering\arraybackslash}p{0.1\linewidth}
        >{\centering\arraybackslash}p{0.1\linewidth}
        >{\centering\arraybackslash}p{0.04\linewidth}
        >{\centering\arraybackslash}p{0.04\linewidth}
        >{\centering\arraybackslash}p{0.04\linewidth}
        >{\centering\arraybackslash}p{0.04\linewidth}
        >{\centering\arraybackslash}p{0.04\linewidth}
        >{\centering\arraybackslash}p{0.07\linewidth}
    }
\hline
\rowcolor[HTML]{C0C0C0} \textbf{Name} & \textbf{Year} & \textbf{Scale} &
\textbf{P} & \textbf{W} & \textbf{E} &
\textbf{H} & \textbf{QR} & \textbf{EF} \\
\hline
Evasive-PDFMal2022 \cite{issakhani2022pdf} & 2022 & 40{,}282 
& $\checkmark$ & - & - & - & - & YES \\
Hossain et al.  \cite{hossain2024pdf} & 2024 & 15{,}958  
& $\checkmark$ & - & - & - & - & YES \\
EMBER2024 \cite{joyce2025ember2024} & 2025 & 64{,}000 & $\checkmark$ & - & - & - & - & YES \\
Casino et al.\cite{casino2023analysis} & 2023 & 15{,}421 & - & $\checkmark$ & $\checkmark$ & - & - & YES \\
Koutsokostas et al. \cite{koutsokostas2022invoice}& 2022 & 18{,}307 & - & $\checkmark$ & $\checkmark$ & - & - & Yes \\
DikeDataset \cite{parkour2013_contagio_counts} & 2023 & N/A & - & \checkmark & \checkmark & - & - & No \\
Contagio Dump \cite{parkour2013_contagio_counts} & 2013 & 28{,}760 & \checkmark & \checkmark & \checkmark & - & - & No \\
PhreshPhish \cite{dalton2025phreshphish} & 2025 & 408{,}917 & - & - & - & \checkmark & - & Yes \\
Ariyadasa et al. \cite{ariyadasa2021_mendeley_html} & 2021 &
80{,}000 
&- & - & - & \checkmark & - & No \\
Singh et al \cite{singh2020malicious} & 2020 & \~ 1.5M & - & - & - & \checkmark & - & Yes \\
Hess et al. \cite{hess2018malicious} & 2018 & 7,303 
&- & - & - & \checkmark & - & Yes \\
Ariyadasa et al. \cite{ariyadasa2022combining} & 2022 & 50{,}000 & - & - & - & \checkmark & - & Yes \\
Yerima et al. \cite{yerima2022malicious} & 2022 &
40{,}282 
& \checkmark & - & - & - & - & Yes \\
Al-Saedi et al.\cite{falah2021improving} & 2021 &
26{,}000 
&
\checkmark & - & - & - & - & Yes \\
Ruaro et al. \cite{ruaro2022symbexcel} & 2022 &
17{,}000 
&- & - & \checkmark & - & - & Yes \\
Sadiq QR Dataset \cite{Sadiq2023QR} & 2023 &
200{,}000 
&- & - & - & - & \checkmark & No \\
Galadima Dataset \cite{galadima_2025_qr1000} & 2025 & 1{,}000 
&- & - & - & - & \checkmark & No \\
Trad et al. \cite{trad2025detecting} & 2025 &
9{,}987 
&- & - & - & - & \checkmark & No \\
Chen et al \cite{chen2023malicious} & 2023 & - & - & - & - & - & - & Yes \\
Hu et al \cite{hu2023ufadf} & 2023 & - & - & - & - & - & - & Yes \\
\hline
\end{tabular}
\label{Literature_review}
\begin{tablenotes}
          \item \scriptsize \textbf{Note:} P= PDF, W= Word Document,  E= Excel Document, H= HTML, QR= QR code, EF= Extracted Features, N/A= not available
    \end{tablenotes}
\end{table*}


Chen et al. \cite{chen2023malicious} introduce a hybrid machine learning detection approach to detect malicious Office macros by testing on a combination of two publicly available datasets. From these two datasets, 123 features in total are extracted. These features are examined by using different classifiers, including Random Forest, MLP, SVM, and KNN, with Random Forest. The work shows that the trained models on the extracted feature outperform significantly. Similarly, 
Hu et al. \cite{hu2023ufadf} suggested a detection model for Malicious Office Documents by extracting six categories of features. UFADF also extracts comprehensive indicators and then feeds them to traditional classifiers such as Random Forest, SVM, XGBoost, and MLP. The experimental results show that compared to trained a model on individual feature groups combining heterogeneous features improves performance significantly.

phreshphish \cite{dalton2025phreshphish} is a real-world dataset consisting of 371,941 HTML-URL pair samples from phishing and benign websites. The samples are evaluated  with a time-separated test set and various benchmark splits to prevent information leakage while obtaining realistic base rates for comparing linear SVM and shallow FNN and BERT-based encoder performance. The Phishing Websites Dataset (Mendeley Data, 2021) \cite{ariyadasa2021_mendeley_html} provides a large corpus of 80k webpages,  where each instance includes the URL and the corresponding HTML page; an index.sql file maps URLs to saved HTML filenames. The HTML content of the webpage serves as practical ground truth for training and testing page-content–based detectors. Singh et al. \cite{singh2020malicious} introduces a new dataset that contains ~1.5M webpages, that serves as a dataset description instead of a modeling benchmark because it explains data acquisition methods and labeling procedures and shows exploratory results but it does not train classifiers and evaluation focuses on label verification and feature distribution inspection. 

The research by Hess et al. \cite{hess2018malicious} introduces a detection model for harmful HTML files through static analysis when dealing with unbalanced and noisy data sets. This work presents 32 static HTML features to test detection and classification algorithms. The results demonstrate that boosting and bagging ensemble methods achieve high accuracy. Similarly, 
Ariyadasa et al. \cite{ariyadasa2022combining} developed a hybrid phishing detection model which utilized Long-Term Recurrent Convolutional Networks (LRCN) and Graph Convolutional Networks (GCN) to analyze both sequential HTML/URL features and hyperlink graph structures. The combined approach enhances resistance to obfuscation methods through hidden forms, malicious scripts and URL manipulation while achieving the best results on benchmark datasets.

Yerima et al. \cite{yerima2022malicious} proposed an enhanced machine learning–based approach model for detecting malicious PDF files by providing 35 features. The 6 anomaly\_based features are designed for capturing deviations such as mismatched object or stream tags and suspicious combination of JavaScript and embedded files. Some classifiers like Random Forest, SVM, and MLP, were used to evaluate the extracted features. Similarly, 
Al-Saedi et al. \cite{falah2021improving} introduced a machine learning model for detecting malicious PDF files by analyzing 13,000 malicious and  13,000 benign PDF files to extract 141 features. The set of features are evaluated by using Random Forest, Decision Tree, SVM and  k-NN classifiers. Then a feature selection techniques is used to discover the most important attributes that simplified the model structure without compromising its accuracy. Another 
research by Ruaro et al. \cite{ruaro2022symbexcel} introduces a static detection framework for malicious Excel  macros called EXCELHunter. The system retrieves multiple features from AST representations to detect both obfuscated and dynamically generated macro code. The research evaluates EXCELHunter through analysis of more than 17,000 benign and malicious Excel samples obtained from VirusTotal and open repositories.

A recent contribution of the Benign and Malicious QR Codes Dataset is published by Sadiq \cite{Sadiq2023QR}. The source dataset includes more than 600,000 URLs; however, only the first 100,000 benign and 100,000 malicious URLs were selected for the QR code generation step. The dataset contains one of the biggest publicly accessible collections for studying Quishing attacks which enables researchers to test image-based detection methods. Another dataset, 1,000 Images of Malicious and Benign QR Codes, was released in 2025 by Galadima \cite{galadima_2025_qr1000}. The dataset provides suitable conditions for binary classification, computer vision–based phishing detection and security analysis of QR code threats because of its balanced structure and controlled variation in URL formats and error-correction levels and encoding densities. A more recent study by Trad et al. \cite{trad2025detecting} introduced a Quishing dataset by selecting 10,000 URLs from PhishStorm and rendering them as QR images in Python. In order to evaluate the generated QR codes, some traditional models such as Logistic Regression, Decision Tree, Naïve Bayes and so on, on the QR-code dataset are used. For this purpose, they flattened each image into a vector and then fed those pixel features to classic ML models.

Despite progress in this area, existing datasets remain fragmented and mostly focus on limited file types, which restrict their ability to represent the diversity of malicious attachments found in real-world emails. To the best of our knowledge, no publicly available dataset simultaneously covers Word, Excel, PDF, HTML, and QR-based samples. To address this gap, this paper introduces a unified dataset containing both benign and malicious samples across these five formats. Table \ref{Literature_review} summarizes the reviewed datasets and studies, noting that while some works (e.g., Chen et al. \cite{chen2023malicious}, Hu et al. \cite{hu2023ufadf}) focus on feature extraction, none provide a dataset as comprehensive and diverse as the one proposed in this paper.

\section{Data Collection}
\label{Nazgol_Malware_methodology_Nazgol_new}
\begin{figure*}[!h]
\begin{center}
\centering
{\includegraphics[width=6 in]{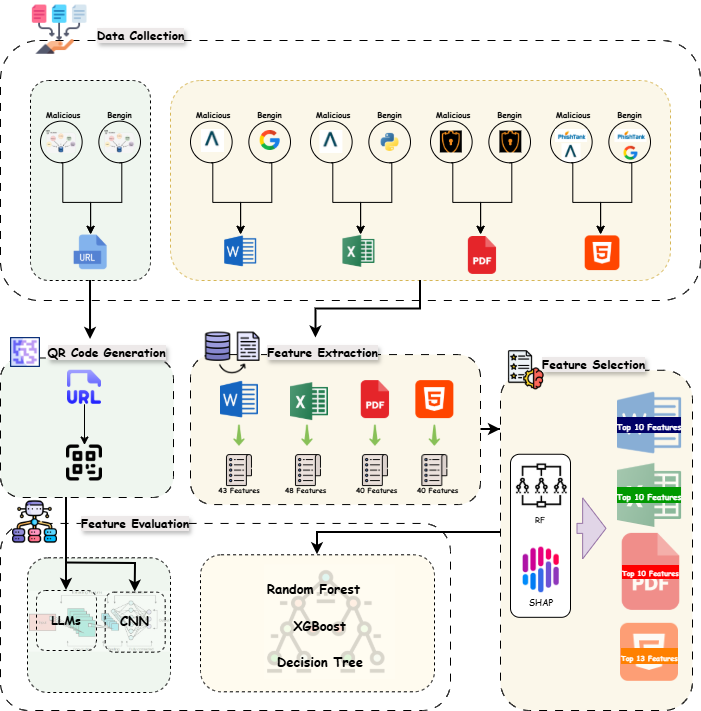}}\\
\caption{Overview of the dataset generation pipeline.}
\label{DataGeneration_Pipeline}
\end{center}
\end{figure*}

In this section, we describe a standardized pipeline for dataset generation, including data collection, feature extraction, feature selection by considering five attachment types (Word document, Excel, PDF, HTML and QR code). Figure \ref{DataGeneration_Pipeline} illustrates the complete dataset generation pipeline that is used in this paper. The first and the most important step is the collection phase, which must be conducted meticulously to ensure dataset integrity. We developed a comprehensive and diverse malicious attachment dataset by collecting samples from five common file formats including Word document, Excel document, PDF, HTML and QR code. To ensure high dataset quality, we carefully selected the sources for each class and file type, as described in the following paragraphs.


\subsection{Word Document}
\begin{itemize}
\item \textbf{Benign Samples:} The collection of benign Word document samples was done by crawling trusted sources such as Google and Wikipedia. The selection of topics was done randomly to achieve diverse writing styles and document structures. The crawled content was saved in \textit{.docx} format to be prepared for feature extraction process. In total, we collected 10,000 legitimate Word document samples. 
\item \textbf{Malicious Samples:} The trusted open-source threat intelligence platform MalwareBazaar \cite{malwarebazaar}, provided access to all updated malicious Word document samples. Approximately 50,000 malicious samples were downloaded, and 10,000 representative samples were selected for static analysis and feature extraction.
\end{itemize}

\subsection{Excel Document}
\begin{itemize}
\item \textbf{Benign Samples:} The collection of 10,000 benign Excel files was done synthetically, as collecting a large number of genuine files was challenging. We generated these files by using a script that choose topics and values randomly. The system generated 10,000 synthetic files to mimic actual business and academic use cases. The generated files were stored in \textit{.xlsx} format and prepared for the subsequent feature extraction phase.
\item \textbf{Malicious Samples:} Malicious Excel samples were collected from MalwareBazaar \cite{malwarebazaar} similar to the procedure used for Word documents. By crawling the repository, approximately 50,000 samples were gathered, from which 10,000 representative samples were selected for further analysis.
\end{itemize}

\subsection{PDF}
\begin{itemize}
\item \textbf{Benign Samples:} The PDFMal2022 dataset \cite{issakhani2022pdf} provided a collection of clean and verified PDF files, which were used as benign samples. A total of 10,000 benign PDF samples were collected to maintain balance between benign and malicious classes.

\item \textbf{Malicious Samples:} Malicious PDF samples were collected from the PDFMal2022 dataset \cite{issakhani2022pdf}, 
generated by the Canadian Institute for Cybersecurity (CIC). This dataset serves as a primary benchmark source for both malicious and benign PDF files. The samples in this dataset were collected from multiple repositories, including Contagio and VirusTotal, and contains a wide range of malware-infected PDFs. From this dataset, 10,000 malicious samples were selected for analysis in this study.

\end{itemize}

\subsection{HTML}
\begin{itemize}
\item \textbf{Benign Samples:} The collecting of benign HTML files was done using two resources. First, legitimate web pages were crawled directly from Google to capture real and diverse HTML structures. Second, the PhishTank dataset \cite{phishtank2024} was used, which provides verified benign and malicious webpage samples. From these two sources, 10,000 legitimate HTML files were selected to ensure dataset diversity and represent real-world webpage characteristics.

\item \textbf{Malicious Samples:} The collection of malicious HTML files was done using the PhishTank dataset \cite{phishtank2024}, which maintains an updated repository of verified phishing webpages submitted by the security community. A total of 10,000 malicious HTML samples were selected from this dataset for further analysis and feature extraction. 
\end{itemize}

\subsection{QR code} 
\begin{figure}[t]
  \centering
  \includegraphics[width=0.8\columnwidth]{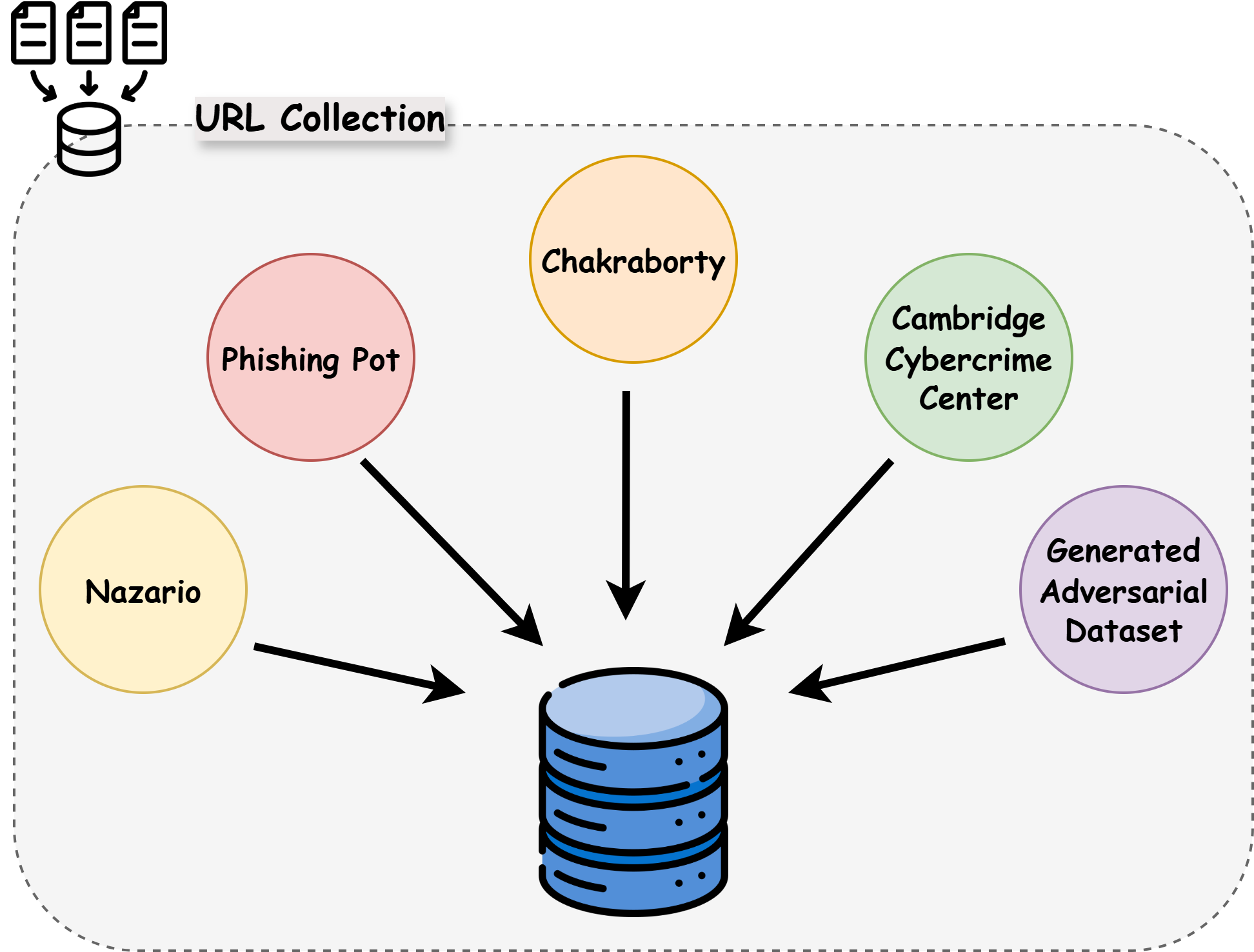}
  \caption{Sources of collected phishing URLs.}
  \label{fig:URLCollection}
\end{figure}

For collecting QR code samples, we gathered one million malicious and benign URLs and generated corresponding QR code images using a Python script. The URL samples were obtained from multiple reliable sources, including publicly available datasets that contain benign and malicious emails with embedded URLs. Figure \ref{fig:URLCollection} illustrates the diverse sources used for collecting the URL data.

\begin{figure}[t]
  \centering
  \begin{minipage}{0.44\linewidth}   
    \centering
    \includegraphics[width=0.9\linewidth]{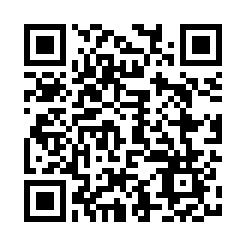}\\ 
    {\footnotesize (a) Benign QR code}
  \end{minipage}
  \hspace{0.01\linewidth}            
  \begin{minipage}{0.44\linewidth}
    \centering
    \includegraphics[width=0.9\linewidth]{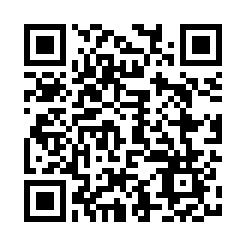}\\
    {\footnotesize (b) Malicious QR code}
  \end{minipage}
  \caption{Examples of benign and malicious QR codes.}
  \label{fig:QR_examples}
\end{figure}

Figure \ref{fig:QR_examples} presents one benign (a) and one malicious (b) QR code sample generated during the collection phase. The two images exhibit identical visual characteristics, making it difficult to distinguish between malicious and benign QR codes based only on their design patterns. The benign QR code corresponds to a legitimate domain, whereas the malicious one encodes a phishing URL that redirects users to a harmful website.

The source of benign and malicious URL samples are provided as follows: 

\begin{itemize}
\item \textbf{Benign Samples} The benign samples were collected from two sources, the Chakraborty Phishing dataset \cite{subhajournal_phishingemails} and the PhiUSIIL Phishing URL \cite{prasad2024phiusiil}, both of which contain verified benign and malicious samples.

\item \textbf{Malicious Samples} The malicious samples were collected from multiple sources, including the Nazario Phishing Email Corpus \cite{gonzalez2011phishing}, Chakraborty Phishing dataset \cite{subhajournal_phishingemails}, Phishing Pot \cite{phishing_pot} , PhiUSIIL Phishing URL\cite{prasad2024phiusiil}, Cambridge Cybercrime Center \cite{cambridgecybercrime2025}, and a generated adversarial dataset created for this study.
\end{itemize}

In summary, Table \ref{tab:dataset-size} presents the breakdown of the dataset by file type.

\begin{table}[t]
\scriptsize
\centering
\caption{Distribution of benign and malicious samples across five file types.}
\label{tab:dataset-size}
\renewcommand{\arraystretch}{1.5}
\begin{tabular}{
        >{\centering\arraybackslash}p{0.25\linewidth}
        >{\centering\arraybackslash}p{0.2\linewidth}
        >{\centering\arraybackslash}p{0.2\linewidth}
        >{\centering\arraybackslash}p{0.2\linewidth}
        } \hline
\rowcolor[HTML]{C0C0C0} Type & Benign & Malicious & Total \\
\hline
DOCX & 10{,}000 & 10{,}000 & 20{,}000 \\
\rowcolor[HTML]{F0F0F0} XLSX & 10{,}000 & 10{,}000 & 20{,}000 \\
PDF  & 10{,}000 & 10{,}000 & 20{,}000 \\
\rowcolor[HTML]{F0F0F0} HTML & 10{,}000 & 10{,}000 & 20{,}000 \\
QR Code  & 430{,}000 & 575{,}000 & 1{,}005{,}000 \\
\bottomrule
\end{tabular}
\end{table}

\subsection{Benign vs. Malicious Sample Acquisition}
The evaluation of machine learning models for malicious attachment detection requires a distinct separation between benign and malicious samples through both content and acquisition methods. Our dataset construction included specific methods to separate the classes while preventing data leakage and artificial similarity and overlapping structures between the two categories.

\textbf{Malicious samples:} The data collection focused exclusively on trusted threat intelligence and malware distribution platforms.  All datasets that are used in this paper provide verified malicious files through community-based analysis or malware sandboxes and incident reports.The selected malicious files often exhibit obfuscation techniques, embedded scripts, macros, or other suspicious artifacts common in real-world attacks. All malicious samples were verified to be intact and suitable for static feature extraction.

\textbf{Benign samples:} In contrast, benign data contained either synthetic content or crawled information from public resources. The benign Word document and HTML files were obtained through Wikipedia and Google content crawling to achieve diversity and document organization and domain variety. The benign Excel files were artificially produced to duplicate typical business and academic spreadsheet formats and value patterns. The benign PDF files were extracted from the PDFMal2022 dataset which contains documents that have been verified as malware-free. Benign URLs that are used for generating QR codes, are extracted from benign emails from different resources.


\section{Feature Engineering and Classification}

Our study focuses on five data types, including Word documents, Excel spreadsheets, PDF files, HTML pages, and QR codes. For the first four file formats, we developed a static feature-based approach, which includes the feature extraction and feature selection phases, followed by the use of traditional machine learning models for the detection phase. This approach enables the development of lightweight and efficient detection models, as it does not rely on dynamic file execution. In contrast, QR codes were analyzed directly using a Convolutional Neural Network (CNN), enabling image-based learning without explicit feature extraction. To further enhance detection performance, a URL lexical analysis was introduced using lightweight transformer models after decoding the URLs embedded within the QR codes.

The development of effective machine learning models for detecting malicious attachments across four data types requires a dedicated feature extraction pipeline capable of capturing both general and format-specific characteristics. The objective was to extract structural properties, embedded content, and behavioral patterns through static analysis methods that do not require file execution. A customized parser was implemented for each document type to extract diverse features, including file metadata, macro indicators, script patterns, embedded object structures, and related attributes.

After feature extraction, the set of extracted features went through a two-step feature selection process to identify the most important attributes for classification tasks and also reduce dimensionality in order to have more efficient training. The SHAP (SHapley Additive exPlanations) model was employed to quantify the contribution of each feature to the model’s prediction outcomes, providing interpretability through feature impact analysis. In parallel, Random Forest feature importance scores were computed to rank attributes based on their contribution to decision splits within the ensemble. The integration of these two evaluation methods allowed us to choose a smaller set of essential features for each file type. The selection process maintains only the most important features which leads to improved detection results and better model generalization capabilities.

\subsection{Word}

In order to extract high level features from Word documents, we created a static feature extraction pipeline which focuses on both high-level and structure-specific characteristics of \textit{.docx} files. The detection system was designed to identify essential malicious indicators, including embedded macros, suspicious VBA keywords, Dynamic Data Exchange (DDE) links, and abnormal XML structures that appear in harmful Word document attachments.
    
The extraction process produces 43 features for each file. The extracted features consist of five main categories which include basic metadata, macro-based content features, DDE patterns, OLE object features, and XML structural features. The system aims to detect multiple malicious document characteristics through static analysis while maintaining lightweight functionality. The key components of the feature extraction phase are as follows:

\begin{itemize}
\item \textbf{Entropy \& Size:} Measures the level of randomness and the overall document size, often indicating embedded payloads or obfuscation.

\item\textbf{Macros \& VBA Keywords:} These two components detect the presence of macros and extracts suspicious script keywords such as AutoOpen, Shell, CreateObject, and PowerShell.

\item\textbf{DDE Detection:} this component identifies the present of DDE patterns, which are known to execute external commands when the document is opened.

\item \textbf{XML Structure Paths \& Attributes:} These components consider for Extracting the number of structural patterns and attributes in the internal XML trees (DOCX files are ZIP archives of XML data).

\item \textbf{OLE Object Indicators:} These features Count the embedded OLE objects and their types.

\end{itemize}

All extracted features are listed in Table \ref{table:word_selected_features_43}.

\begin{table*}[ht]  
\centering
\scriptsize
\caption{Summary of extracted features for Word document analysis.} 
\label{table:word_selected_features_43} 
\renewcommand{\arraystretch}{1.5}
\begin{tabular}{>{\arraybackslash}p{0.25\linewidth}
        >{\arraybackslash}p{0.57\linewidth}
        >{\arraybackslash}p{0.05\linewidth}}
 \bottomrule 
\rowcolor[HTML]{C0C0C0} \textbf{Feature Group} & \textbf{Purpose (Brief Description)} & \textbf{No.} \\ 
\hline
Basic Metadata & Size, entropy (obfuscation signal) & 2 \\ 
\rowcolor[HTML]{F0F0F0} Macro / VBA Indicators & Macro presence, suspicious VBA keywords & 2 \\ 
DDE Detection & DDE patterns that can trigger external execution & 1 \\ 
\rowcolor[HTML]{F0F0F0}OLE Object Indicators & Embedded OLE objects (counts/types) & 2 \\ 
XML Structure Paths \& Attributes & WordprocessingML structural attributes and path frequencies& 36 \\ \hline
\rowcolor[HTML]{F0F0F0} \textbf{Total} & & \textbf{43} \\ 
\hline
\end{tabular} 
\begin{tablenotes}
          \item \scriptsize \textbf{Note:} No= Number of extracted Features
    \end{tablenotes}
\end{table*}

In the feature extraction phase, 43 static features were extracted from Word documents; however, not all of them were useful for classification task. Therefore, we applied the two-step feature selection process described earlier, using SHAP and Random Forest feature importance analysis to identify the most informative and lightweight features. The final set of 10 features was selected based on their high importance rankings in both SHAP and Random Forest evaluation. These features combine strong predictive capabilities with efficient computation which makes them appropriate for real-time malicious document detection systems. The final set of selected features for Word documents is listed in Table \ref{table:word_top_features}. 

\begin{table*}[ht]
\scriptsize
\centering
\caption{Top 10 Selected Features for Word Document Using SHAP \& Feature Importance.}
\label{table:word_top_features}
\renewcommand{\arraystretch}{1.5}
\begin{tabular}{
        >{\arraybackslash}p{0.03\linewidth}
        >{\arraybackslash}p{0.2\linewidth}
        >{\arraybackslash}p{0.6\linewidth}
                } \bottomrule 
\rowcolor[HTML]{C0C0C0} \textbf{No.} & \textbf{Feature Name} & \textbf{Description} \\
\hline
1  & ole\_object\_count       & Number of embedded OLE objects in the document \\
\rowcolor[HTML]{F0F0F0} 2  & ole\_object\_type\_count & Number of distinct types of embedded OLE objects \\
3  & macro\_present           & Binary flag indicating the presence of embedded VBA macros \\
\rowcolor[HTML]{F0F0F0} 4  & dde\_present             & Detects presence of DDE command patterns that can trigger external process execution \\
5  & vba\_keywords\_count     & Number of suspicious VBA macro keywords (e.g., \texttt{createobject}, \texttt{powershell}) \\
\rowcolor[HTML]{F0F0F0} 6  & entropy                  & Shannon entropy of the document's raw content, indicating randomness or obfuscation \\
7  & struct\_ContentType      & Frequency of \texttt{ContentType} attributes in the DOCX internal XML \\
\rowcolor[HTML]{F0F0F0} 8  & struct\_PartName         & Count of structural \texttt{PartName} references used in the document package \\
9  & file\_size               & Size of the Word document in bytes \\
\rowcolor[HTML]{F0F0F0} 10 & struct\_pos              & Frequency of \texttt{pos} attributes in XML structure (layout positioning) \\
\hline
\end{tabular}
\end{table*}



\subsection{Excel}

In order to analyze Excel attachment files statically, we created a complete feature extraction pipeline. This system identifies structural, content-based, and macro-related attributes associated with obfuscation and suspicious behavior in Excel documents. The extraction framework extracts 48 unique features from each file across five high-level categories, including file-level metadata, sheet/cell structure, cell content statistics, macro code metrics, and embedded resource/external interaction features.

In addition, the system uses Optical Character Recognition (OCR) to extract textual information from preview images embedded in Excel files which helps reveal hidden deceptive content. We excluded preview image text that embeds crawl hints (e.g., site watermarks).

The main categories of the extracted features are summarized as follows:

\begin{itemize}
\item \textbf{Entropy \& Size:} This group of features measures overall scale and text randomness that may indicate embedded payloads or obfuscation.

\item \textbf{Cell Content \& Encoding:} These features identify  what is inside cells and flags encoded data.

\item \textbf{Macro Behavior \& Code Metrics:} This category captures VBA presence and obfuscation complexity in Excel files.

\item \textbf{External References \& System Interaction:} This set of features indicates potential data exfiltration or code loading by using external links/APIs.

\item \textbf{Structure \& OCR Signals:} The last category reveals any hidden components via layout and image-based lures.
\end{itemize}


\begin{table*}[ht]
\scriptsize
\centering
\caption{Summary of Extracted Features for Excel Document Analysis.}
\label{table:excel_extracted_features}
\renewcommand{\arraystretch}{1.5}
\begin{tabular}{
        >{\arraybackslash}p{0.27\linewidth}
        >{\arraybackslash}p{0.6\linewidth}
        >{\arraybackslash}p{0.04\linewidth}
                } \bottomrule 
\rowcolor[HTML]{C0C0C0} \textbf{Feature Group} & \textbf{Purpose (Brief Description)} & \textbf{No.} \\
\hline
Basic Metadata            & Captures general file and worksheet dimensions                          & 5  \\
\rowcolor[HTML]{F0F0F0} Cell Content              & Describes the content characteristics of spreadsheet cells              & 6  \\
Formula and Hyperlinks    & Indicates spreadsheet complexity and external linking                   & 2  \\
\rowcolor[HTML]{F0F0F0} Macro Structure           & Describes macro logic structure and complexity                          & 11 \\
Advanced Macro Features   & Captures deeper syntax patterns                                         & 7  \\
\rowcolor[HTML]{F0F0F0} Behavioral Indicators     & Detects interaction with system-level resources or external references  & 5  \\
Sheet Properties          & Describes worksheet configurations                                      & 6  \\
\rowcolor[HTML]{F0F0F0} Preview Image             & Leverages embedded images for content-based cues                        & 5  \\
\hline
\textbf{Total}            &                                                                        & \textbf{48} \\
\bottomrule
\end{tabular}
\begin{tablenotes}
          \item \scriptsize \textbf{Note:} No= Number of extracted Features
    \end{tablenotes}
\end{table*}

All 48 extracted features from Excel files are summarized in Table \ref{table:excel_extracted_features}. 

Although the initial set of 48 features was carefully extracted based on multiple factors, not all of them contributed equally to the classification model. Similar to the feature selection techniques applied for the previous file formats, we selected the top 10 most effective features that play a critical role in the detection phase. These selected features from Excel files are summarized in Table \ref{table:excel_top_features}.


\begin{table*}[ht]
\scriptsize
\centering
\caption{Top 10 Selected Features for Excel Document Using SHAP \& Feature Importance.}
\renewcommand{\arraystretch}{1.5}
\label{table:excel_top_features}
\begin{tabular}{
        >{\arraybackslash}p{0.02\linewidth}
        >{\arraybackslash}p{0.2\linewidth}
        >{\arraybackslash}p{0.65\linewidth}
                } \bottomrule 
\rowcolor[HTML]{C0C0C0} \textbf{No.} & \textbf{Feature Name} & \textbf{Description} \\
\hline
1  & entropy\_of\_text                 & Shannon entropy of extracted worksheet text; higher values suggest randomness/obfuscation or encoded payloads. \\
\rowcolor[HTML]{F0F0F0} 2  & macro\_chr\_count                 & Number of \texttt{Chr}/\texttt{ChrW}-style function uses in VBA, often used to construct strings and evade signatures. \\
3  & macro\_vocab\_size                & Count of unique tokens in VBA macros; a proxy for macro code diversity/complexity. \\
\rowcolor[HTML]{F0F0F0} 4  & macro\_arithmetic\_operator\_count& Number of arithmetic operators (e.g., \texttt{+}, \texttt{-}, \texttt{*}, \texttt{/}, \texttt{Mod}) used in macros; indicative of obfuscation logic. \\
5  & macro\_token\_count               & Total token count in VBA macros; measures overall macro size/complexity. \\
\rowcolor[HTML]{F0F0F0} 6  & macro\_max\_line\_length          & Length of the longest line in VBA code; unusually long lines can hide concatenated or obfuscated payloads. \\
7  & remote\_template\_present         & Binary flag indicating a reference to a remote template (external URL), which can enable external code loading. \\
\rowcolor[HTML]{F0F0F0} 8  & numeric\_cell\_count              & Number of cells with numeric values across worksheets; helps profile content composition. \\
9  & string\_cell\_count               & Number of cells with string values; large counts can correlate with embedded text or lures. \\
\rowcolor[HTML]{F0F0F0} 10 & avg\_cell\_length                 & Average character length of string cell contents; higher values may indicate embedded/encoded data. \\
\hline
\end{tabular}
\end{table*}


\subsection{PDF}

To analyze PDF attachments statically, we designed a feature extraction pipeline focused on capturing low-level structural, behavioral, and metadata attributes. The system aimed to detect concealed malicious indicators which include hidden JavaScript code, exploitative actions, and obfuscated content that malicious PDFs typically use.

The framework extracts 40 features from each PDF file. These features are categorized into six main groups which include metadata \& object statistics, structural \& forms, stream \& entropy analysis, and scripted actions \& external references (JavaScript and URI). The system uses PyMuPDF, PyPDF2, and pdfminer, with an OCR-based fallback, in order to analyze different PDF types, including encrypted and image-only PDFs. The extracted feature categories are described as follows:

\begin{itemize}

\item \textbf {Stream Analysis and Entropy:} These components calculate stream entropy and average stream size that appear between stream and end stream markers. The high entropy levels or large stream sizes show possible payload obfuscation. 

\item \textbf {JavaScript and Actions:} The tool measures the number of occurrences for JavaScript, Launch, OpenAction, SubmitForm and URI entries (per-type counts) to identify auto-execution vectors which activate when files open or users interact with them.

\item \textbf{External references \& networking:} The feature identifies any URL or action that uses non-standard ports which do not match 80 or 443. The function returns 1 when such a port exists but returns 0 when it does not.
    
\item \textbf {Metadata \& Object Statistics:} The tool calculates file size information along with encryption flag status, object count, font object count and page number data as these metrics often correlate with suspicious manipulation or content embedding.

\item \textbf {Obfuscation Indicators}: These features evaluate name\_obfuscations through encoded object names and searches for \/Filter[(nested filters) which might conceal payloads.

\item \textbf {Fallback Text Recovery:} It employs pdfminer, pdftotext and OCR as final extraction methods in order to achieve the highest possible document interpretability. The recorded text length serves as an indicator for both visibility and content density.

\item \textbf{Structural \& Forms:} These features captured embedded content and interactive form structures that is able to be delivered with  payload or user-driven execution paths.

\end{itemize}

\begin{table*}[ht]
\scriptsize
\centering
\caption{Summary of Extracted Features for PDF File Analysis.}
\label{table:pdf_selected_features_40}
\renewcommand{\arraystretch}{1.5}
\begin{tabular}{
        >{\arraybackslash}p{0.31\linewidth}
        >{\arraybackslash}p{0.59\linewidth}
        >{\arraybackslash}p{0.03\linewidth}
        }
\bottomrule
\rowcolor[HTML]{C0C0C0} \textbf{Feature Group} & \textbf{Purpose (Brief Description)} & \textbf{No.} \\
\hline
Basic Metadata         & Size, number of pages, encryption status, and metadata footprint                       & 4 \\
\rowcolor[HTML]{F0F0F0} Content Properties     & Readable text and title length, embedded images, and OCR fallback signal         & 4 \\
Stream Analysis        & Stream usage statistics, size, entropy, and number of object streams                    & 5 \\
\rowcolor[HTML]{F0F0F0} Object Statistics     & Counts of objects, fonts, and cross-reference tables or entries                   & 4 \\
Embedded Files         & Presence and typical size of embedded attachments                                       & 2 \\
\rowcolor[HTML]{F0F0F0} Obfuscation Indicators & Encoded names and nested filters that may conceal payloads                        & 2 \\
JavaScript / URI Indicators & Script references and presence of external link indicators                          & 3 \\
\rowcolor[HTML]{F0F0F0} Action Triggers        & Auto or interactive actions (e.g., \texttt{Launch}, \texttt{OpenAction}, \texttt{AA}) & 5 \\
Form Features          & Interactive form structures such as AcroForm or XFA                                    & 2 \\
\rowcolor[HTML]{F0F0F0} Encoding Techniques \& Media & Advanced codecs, compression, or media filters (e.g., JBIG2, RichMedia)          & 4 \\
Behavioral Correlation & Co-occurrence of multiple risky behaviors within the same object                        & 1 \\
\rowcolor[HTML]{F0F0F0} Structure Validity     & Header validity and key trailer/startxref marker consistency                      & 3 \\
Non-Standard Usage      & Presence of URLs or actions using non-standard network ports                           & 1 \\
\midrule
\textbf{Total}          &                                                                       & \textbf{40} \\
\bottomrule
\end{tabular}
\begin{tablenotes}
    \item \scriptsize \textbf{Note:} No = Number of Extracted Features
\end{tablenotes}
\end{table*}


Similar to the feature extraction procedures applied to the previous file formats, 40 features were extracted from PDF files, followed by the selection of the top ten features that demonstrated the highest predictive impact during evaluation. These selected features emphasize metadata footprint, structural richness, stream activity/entropy, and header validity-signals that proved most discriminative in our evaluations. The complete set of extracted features is summarized in Table~\ref{table:pdf_selected_features_40}, while the selected top features for PDF files are listed in Table~\ref{table:pdf_selected_features}.


\begin{table*}[ht]
\scriptsize
\centering
\caption{Top 10 Selected Features for PDF Document Using SHAP \& Feature Importance.}
\label{table:pdf_selected_features}
\renewcommand{\arraystretch}{1.5}
\begin{tabular}{
        >{\arraybackslash}p{0.03\linewidth}
        >{\arraybackslash}p{0.12\linewidth}
        >{\arraybackslash}p{0.7\linewidth}
                } \bottomrule 
\rowcolor[HTML]{C0C0C0} \textbf{No.} & \textbf{Feature Name} & \textbf{Description} \\
\hline
1  & text\_length        & Total number of text characters extracted; very low values can indicate image-only or obfuscated text. \\
\rowcolor[HTML]{F0F0F0} 2  & total\_filters      & Count of filter operators referenced in streams (e.g., \texttt{/FlateDecode}, \texttt{/LZWDecode}); proxies encoding complexity. \\
3  & title\_chars        & Number of characters in the document’s \texttt{/Title} metadata field. \\
\rowcolor[HTML]{F0F0F0} 4  & file\_size          & Size of the PDF file in bytes. \\
5  & object\_count       & Total number of indirect objects defined in the document. \\
\rowcolor[HTML]{F0F0F0} 6  & stream\_count       & Number of stream objects (binary data sections). \\
7  & endstream\_count    & Count of \texttt{endstream} markers closing stream objects (consistency with \texttt{stream\_count}). \\
\rowcolor[HTML]{F0F0F0} 8  & metadata\_size      & Size in bytes of the \texttt{/Metadata} (XMP) stream; 0 if absent. \\
9  & valid\_pdf\_header  & Binary flag indicating a valid \texttt{\%PDF-} header with version token. \\
\rowcolor[HTML]{F0F0F0} 10 & entropy\_of\_streams& Average Shannon entropy of stream contents; higher values suggest compression/encryption/obfuscation. \\
\hline
\end{tabular}
\end{table*}

\subsection{HTML}

The extraction of efficient attributes from HTML files requires a complete static analysis pipeline that considers both structural and content-level features. The proposed extraction system analyzes the entire HTML body structure which includes scripts and tags together with attributes and embedded links; therefore, it captures behavioral and obfuscation patterns which are typically used in phishing and other malicious HTML attachments by attacker.

By using this pipeline, 40 features were extracted from each HTML file that cover different behavioral categories such as content entropy, embedded JavaScript, tag composition, URL structure, and obfuscation techniques. The system aims to identify suspicious behaviors through static analysis without relying on dynamic execution. The extraction pipeline consists of the following essential components:

\begin{itemize}
    \item \textbf {Entropy \& Size Indicators:} This set of features focuses on measuring document entropy and script block randomness, which often indicate obfuscation or payload encoding.
    \item \textbf {HTML Tag Composition:} This feature summarizes the total number of tags by considering unique tag types, nesting depth, and embedded components that can be an evidence of anomalies. 
    
    \item \textbf {JavaScript Analysis:} Tracks the frequency of scripts and also the size of embedded JavaScript code as well as the presence of suspicious functions such as \textit{eval()} and redirect patterns like \textit{window.location}.

    \item \textbf {URL \& Link Behavior:} The features in this category considers multiple indicators that include external/internal link counts, base64-encoded segments, hostname digit ratios, average subdomain counts, suspicious URL characters that may indicate redirection or obfuscation.

    \item \textbf {Obfuscation \& Suspicious Indicators:} This set of  features includes hidden iframes, high base64 occurrence count, hex encoding rates and suspicious words such as \textit{login}, \textit{password}, and \textit{secure}. These words are commonly used by attackers in malicious HTML files.
\end{itemize}

All extracted features are summarized in Table \ref{table:html_selected_features_40}. 

\begin{table*}[ht]
\centering
\caption{Summary of Extracted Features for HTML File Analysis.}
\label{table:html_selected_features_40}
\scriptsize
\renewcommand{\arraystretch}{1.5}
\begin{tabular}{p{0.25\linewidth} p{0.60\linewidth} p{0.08\linewidth}}
\hline
\rowcolor[HTML]{C0C0C0} \textbf{Feature Group} & \textbf{Purpose (Brief Description)} & \textbf{No.} \\
\hline
Basic Metadata                     & Size, lines, overall entropy, whitespace formatting levels                    & 5 \\
\rowcolor[HTML]{F0F0F0} Tag Structure      & Tag count/variety, nesting depth, comments, \texttt{noscript}/\texttt{object} usage & 6 \\
JavaScript Indicators             & Script presence/size/complexity, risky functions (e.g., \texttt{eval})        & 7 \\
\rowcolor[HTML]{F0F0F0} Form Features      & HTML forms enabling data capture                                             & 1 \\
Iframe Analysis                   & Presence of iframes, including hidden ones                                    & 2 \\
\rowcolor[HTML]{F0F0F0} Redirection Indicators & Automatic redirection mechanisms                                            & 1 \\
Obfuscation Techniques            & Base64 usage, hex encoding, escaped characters                                & 3 \\
\rowcolor[HTML]{F0F0F0} Suspicious Content Keywords & Phishing keyword frequency and keyword-to-text ratio                          & 2 \\
URL and Link Analysis             & Link structure, lengths, subdomains, digits/punctuation signals               & 11 \\
\rowcolor[HTML]{F0F0F0} Image Analysis      & Image tag usage                                                              & 1 \\
Event-Based Behavior             & Suspicious event handlers (e.g., \texttt{onload}, \texttt{onclick})           & 1 \\
\hline
\rowcolor[HTML]{F0F0F0} \textbf{Total} & & \textbf{40} \\
\bottomrule
\end{tabular}
\end{table*}


Similarly, 40 features were extracted from HTML files, followed by the selection of the top 13 features that demonstrated the highest predictive impact during evaluation. The selected top 13 features for HTML files are presented in Table \ref{table:html_top13_features}.

\begin{table*}[!h]
\centering
\scriptsize
\caption{Top 13 Selected Features for HTML Files Using SHAP \& Feature Importance.}
\label{table:html_top13_features}
\renewcommand{\arraystretch}{1.5}
\begin{tabular}{
        >{\arraybackslash}p{0.03\linewidth}
        >{\arraybackslash}p{0.12\linewidth}
        >{\arraybackslash}p{0.8\linewidth}
                } \bottomrule 
\rowcolor[HTML]{C0C0C0}  \textbf{No.} & \textbf{Feature Name} & \textbf{Description} \\
\hline
1  & url\_punct\_char\_count     & Total punctuation characters across all extracted URLs (e.g., \texttt{/ - = ? \& : . \_}). Higher counts can indicate obfuscation or heavy query strings. \\
\rowcolor[HTML]{F0F0F0} 2  & tag\_count                  & Total number of HTML tags in the document (overall structural volume/complexity). \\
3  & whitespace\_ratio           & Proportion of whitespace to non-whitespace in visible text; unusual spacing can signal templating or padding/obfuscation. \\
\rowcolor[HTML]{F0F0F0} 4  & entropy                     & Shannon entropy of the HTML source; higher values can indicate packed/obfuscated content. \\
5  & form\_count                 & Number of \texttt{<form>} elements; often higher in credential-harvesting pages. \\
\rowcolor[HTML]{F0F0F0} 6  & embedded\_js\_count         & Number of inline/embedded JavaScript blocks (e.g., \texttt{<script>} without external \texttt{src}). \\
7  & html\_whitespace\_ratio     & Whitespace proportion computed over the raw HTML (including markup); complements \texttt{whitespace\_ratio} which targets visible text. \\
\rowcolor[HTML]{F0F0F0} 8  & script\_entropy             & Average Shannon entropy computed over JavaScript blocks; elevated values suggest obfuscated/minified JS. \\
9  & min\_link\_length           & Minimum character length among extracted URLs; extreme short/long values can be indicative of shorteners or crafted links. \\
\rowcolor[HTML]{F0F0F0} 10 & external\_link\_count       & Count of hyperlinks pointing to external hosts (e.g., \texttt{<a href>} with a different hostname). \\
11 & total\_script\_characters   & Total characters across all JavaScript blocks (script volume/complexity proxy). \\
\rowcolor[HTML]{F0F0F0} 12 & internal\_link\_count       & Count of hyperlinks pointing within the same host (internal navigation density). \\
13 & url\_digit\_count           & Total digit characters appearing across all extracted URLs; often higher in crafted/parameterized URLs. \\
\bottomrule
\end{tabular}
\end{table*}

\subsection{Classification Phase}

The features extracted in the previous sections were evaluated separately to determine their effectiveness in distinguishing malicious samples from benign ones across different data types. To perform this evaluation, we used three well-established machine learning classifiers including Random Forest, XGBoost and Decision tree classifiers. The reason for selecting these three models is their interpretability, robustness to feature scaling, and strong performance on structured datasets.

To evaluate both the extracted and selected features, the machine learning models were first trained using the complete feature set to assess overall performance, and then retrained using the selected subset to obtain a representative feature set suitable for lightweight models, making them more appropriate for deployment in resource-constrained environments and email gateways.


\begin{figure*}[t]
\centering \includegraphics[width=7in,height=2 in,keepaspectratio=false]{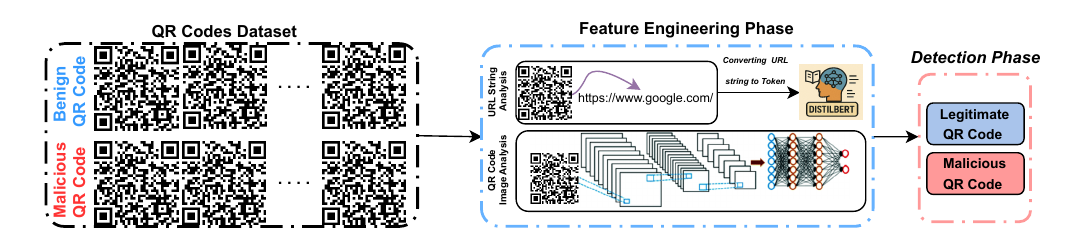}
\caption{Proposed Quishing detection methodology and feature engineering.}
\label{BERT_CNN}
\end{figure*}


QR code classification requires a separate approach, as these data are image-based and therefore demand image processing techniques for both feature extraction and classification. To perform both tasks efficiently, convolutional neural network (CNN) techniques are particularly convenient. Accordingly, a basic CNN-based model was employed to automatically identify discriminative image features for QR code analysis through automated learning instead of requiring human-designed features. After applying the CNN model, we also examined the URLs embedded in the QR codes by using discriminative lightweight language models, including BERT-Tiny, DeBERTa-v3, ModernBERT, and DeepSeek-R1 (Distill-Qwen), to assess their detection performance. The evaluation results are presented in the next section. Figure \ref{BERT_CNN} illustrates the proposed Quishing detection methodology, encompassing two complementary approaches: one based on QR code decoding and URL tokenization, and the other utilizing CNN-driven image analysis.

\section{{Experimental Results and Evaluation}}

\label{Nazgol_Malware/Evaluation_new}
This section provides the procedure used to assess the performance and reliability of the proposed malicious attachment dataset. The main goal of this evaluation is to verify how well the extracted and selected features can distinguish malicious samples from benign ones. All experiments were conducted using standardized evaluation metrics and consistent data-splitting methods for training and testing. The following subsections explains the evaluation metrics, detailed results per file type, and insights into the dataset’s robustness and effectiveness in details.



\subsection{Evaluation Metrics}

The classification performance was evaluated by using standard evaluation metrics such as accuracy, precision, recall and F1-score. Accuracy is a sign how well the model could predict the samples correctly, while precision focuses on the proportion of predicted malicious samples that were actually malicious to reduce false positives. The recall metric evaluates how well a model detects all malicious samples by minimizing false negatives which remains essential for security applications.  Finally, the F1-score is a factor that balances precision and recall by combining both metrics together. It performs well in dealing with imbalanced datasets. These four metrics and their corresponding formula are provided below.

{\small
\begin{equation}
\text{Accuracy} = \frac{TP + TN}{TP + TN + FP + FN}
\end{equation}}

{\small
\begin{equation}
\text{Precision} = \frac{TP}{TP + FP}
\end{equation}}

{\small
\begin{equation}
\text{Recall} = \frac{TP}{TP + FN}
\end{equation}}

{\small
\begin{equation}
F1\text{-score} = 2 \times \frac{\text{Precision} \times \text{Recall}}{\text{Precision} + \text{Recall}}
\end{equation}}

\subsection{Evaluation Results}

In this section, we evaluate the effectiveness of our malicious attachment datasets and QR code samples. For the first four data types including Word doc, Excel, PDF and HTML, we applied three well known machine learning classifiers: Random Forest, XGBoost, and Decision Tree to a balanced dataset of benign and malicious samples containing 20,000 samples (10,000 benign and 10,000 malicious), using a 70/30 split for training and testing, respectively. The second approach evaluates the QR code samples, which is done by applying a CNN model to QR code images, and lightweight language models (LLMs) to analyze the corresponding embedded URLs.

\subsubsection{Word Doc Evaluation}
\begin{figure*}[t]
  \centering
  \begin{minipage}{0.30\textwidth}
    \centering
    \includegraphics[width=\linewidth]{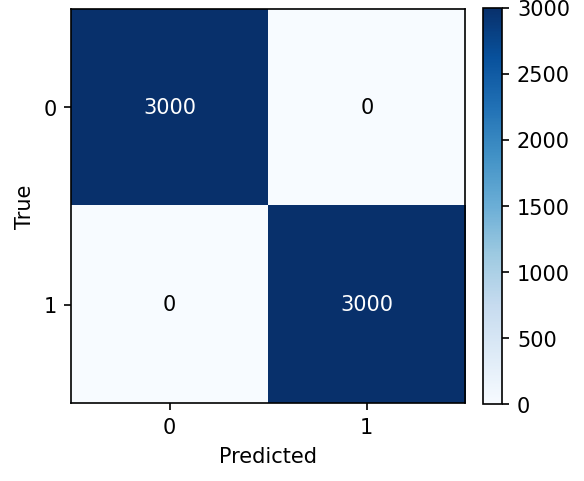}\\
    {\footnotesize (a) Random Forest}
  \end{minipage}\hfill
  \begin{minipage}{0.30\textwidth}
    \centering
    \includegraphics[width=\linewidth]{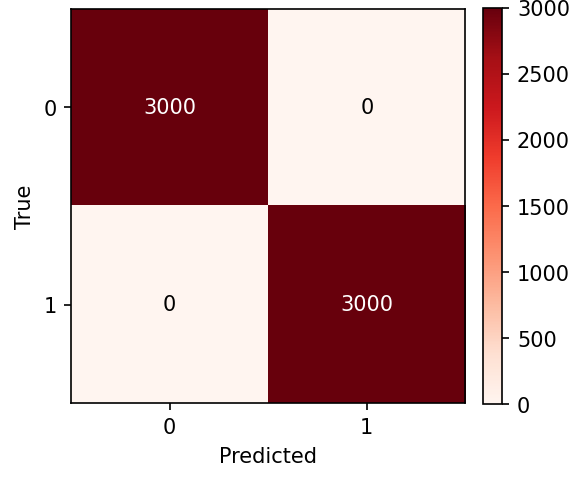}\\
    {\footnotesize (b) XGBoost}
  \end{minipage}\hfill
  \begin{minipage}{0.30\textwidth}
    \centering
    \includegraphics[width=\linewidth]{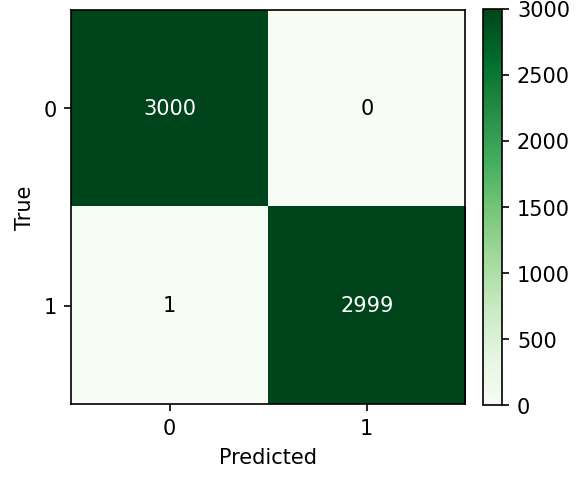}\\
    {\footnotesize (c) Decision Tree}
  \end{minipage}
  \caption{Confusion matrices of Random Forest, XGBoost, and Decision Tree classifiers for Word Document files}
  \label{fig:Doc_ConfusionMatrix}
\end{figure*}

In the feature extraction step, we extracted 43 static features from Word Document files, as described in section 4. To obtain a compact set of efficient and discriminative features, we selected top 10 features from SHAP and Random Forest feature importance rankings. These selected features were evaluated using three classifiers, and the detailed results are presented in Table \ref{Performance_Table}. Additionally,  Figure \ref{fig:url_distribution} (a) provides a visual representation of the results. The selected set of 10 features achieved 100\% Accuracy, Precision, Recall, and F1-score, demonstrating the strong effectiveness of the selected feature subset.
The confusion matrices for all three classifiers are shown in Figure \ref{fig:Doc_ConfusionMatrix}. These matrices confirm the near-perfect classification performance of benign vs. malicious Word Document files by considering the top 10 selected features. Random Forest and XGBoost achieved perfect classification with no misclassifications, while the Decision Tree confusion matrix shows there is only a negligible error rate. These results confirm that the top 10 selected features enable highly accurate and efficient detection of malicious Word document files.




\subsubsection{Excel Evaluation}

\begin{figure*}[t]
  \centering
  \begin{minipage}{0.30\textwidth}
    \centering
    \includegraphics[width=\linewidth]{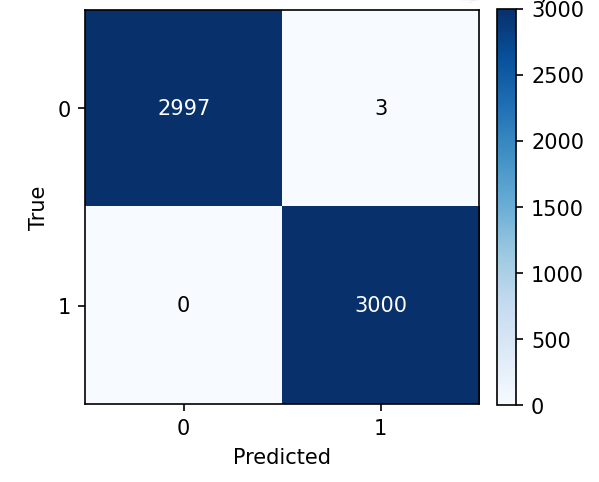}\\
    {\footnotesize (a) Random Forest}
  \end{minipage}\hfill
  \begin{minipage}{0.30\textwidth}
    \centering
    \includegraphics[width=\linewidth]{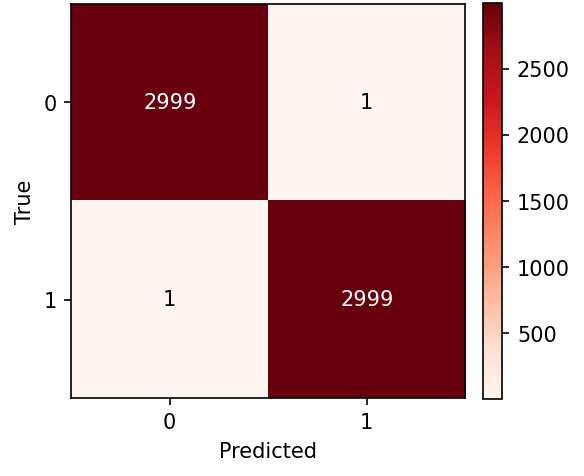}\\
    {\footnotesize (b) XGBoost}
  \end{minipage}\hfill
  \begin{minipage}{0.30\textwidth}
    \centering
    \includegraphics[width=\linewidth]{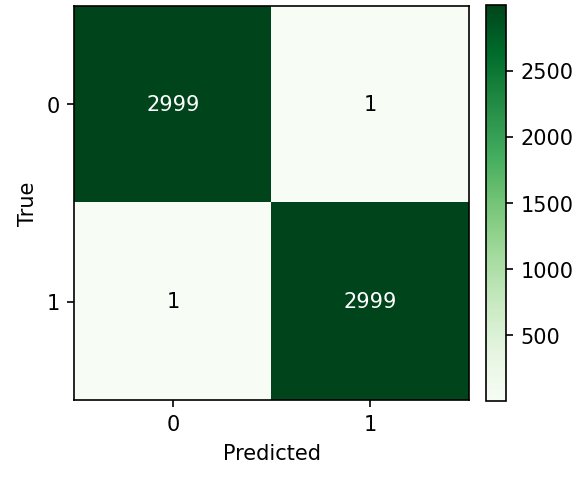}\\
    {\footnotesize (c) Decision Tree}
  \end{minipage}
  \caption{Confusion matrices of Random Forest, XGBoost, and Decision Tree classifiers for Excel Document files}
  \label{fig:Excel_ConfusionMatrix}
\end{figure*}

In order to evaluate the extracted and selected Excel features, we performed the set of experiments that were conducted on Word documents. As explained before, a top 10 features were selected out of 48 extracted features and then the three machine learning classifiers were used for evaluation phase. Figure \ref{fig:url_distribution} (b) illustrates the four evaluation metrics, while Table \ref{Performance_Table} presents their detailed results. Additionally, Figure \ref{fig:Excel_ConfusionMatrix} shows the confusion matrices of all three classifiers as well, which demonstrates the effectiveness of the extracted features.

\subsubsection{PDF Evaluation}

For PDF documents, a total of 40 features were initially extracted, as discussed in Section 4. Subsequently, the top 10 features were selected based on their importance scores. Similar to the previous file formats, these top 10 features were evaluated using the three classifiers described earlier. Table \ref{Performance_Table} presents the detailed evaluation metrics. The results show that the selected features are highly effective, enabling accurate and efficient classification of benign and malicious PDF files. The results of these three models are provided in Figure \ref{fig:url_distribution} (c), which shows that all four evaluation metrics, Accuracy, Precision, Recall, and F1-score, exceed 99\%.

\begin{figure*}[t]
  \centering
  \begin{minipage}{0.30\textwidth}
    \centering
    \includegraphics[width=\linewidth]{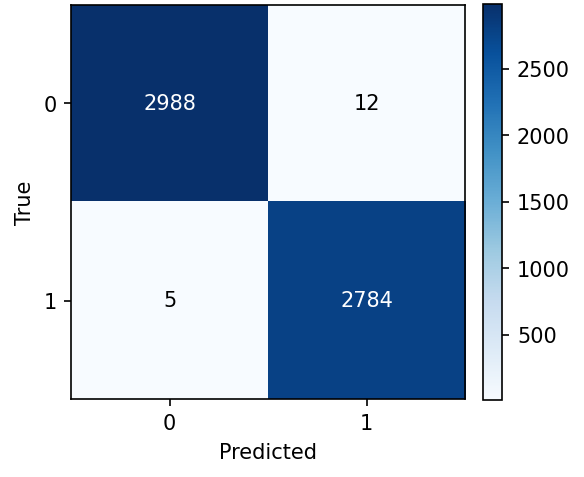}\\
    {\footnotesize (a) Random Forest}
  \end{minipage}\hfill
  \begin{minipage}{0.30\textwidth}
    \centering
    \includegraphics[width=\linewidth]{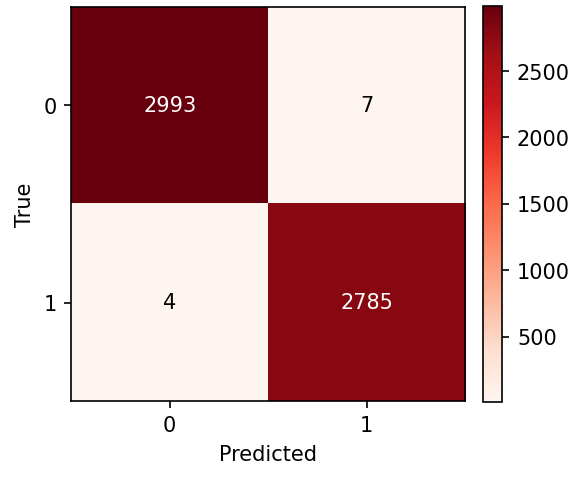}\\
    {\footnotesize (b) XGBoost}
  \end{minipage}\hfill
  \begin{minipage}{0.30\textwidth}
    \centering
    \includegraphics[width=\linewidth]{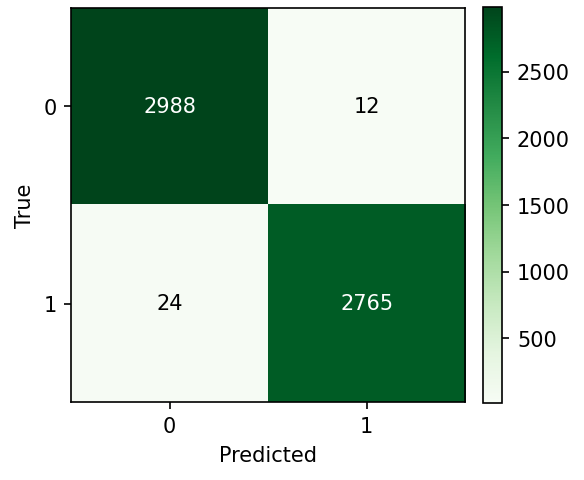}\\
    {\footnotesize (c) Decision Tree}
  \end{minipage}
  \caption{Confusion matrices of Random Forest, XGBoost, and Decision Tree classifiers for PDF files}
  \label{fig:PDF_ConfusionMatrix}
\end{figure*}

The confusion matrices for all three classifiers are shown in Figure \ref{fig:PDF_ConfusionMatrix}. Based on these confusion matrices, the Decision Tree classifier has a negligible misclassification rate on the validation samples, but no misclassification occurred on the test set. Overall, the confusion matrices show that the selected 10 features are highly practical in detecting malicious PDFs with a great accuracy beside improving efficiency.



\subsubsection{HTML Evaluation}

\begin{figure*}[!h]
    \centering
    \begin{subfigure}[b]{0.45\textwidth}
        \centering
        \includegraphics[width=2.5 in]{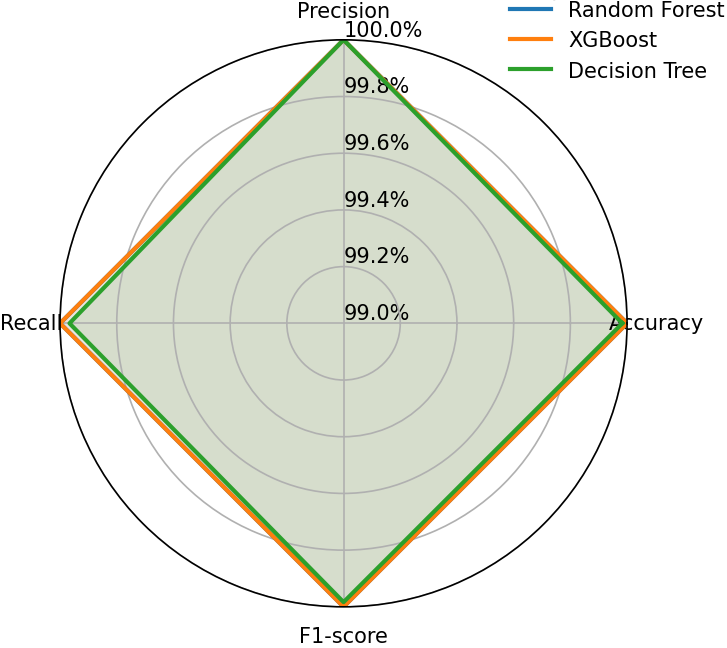}
        \caption{Word Doc}
        \label{????}
    \end{subfigure}
    \hfill
    \begin{subfigure}[b]{0.45\textwidth}
        \centering
        \includegraphics[width=2.5 in]{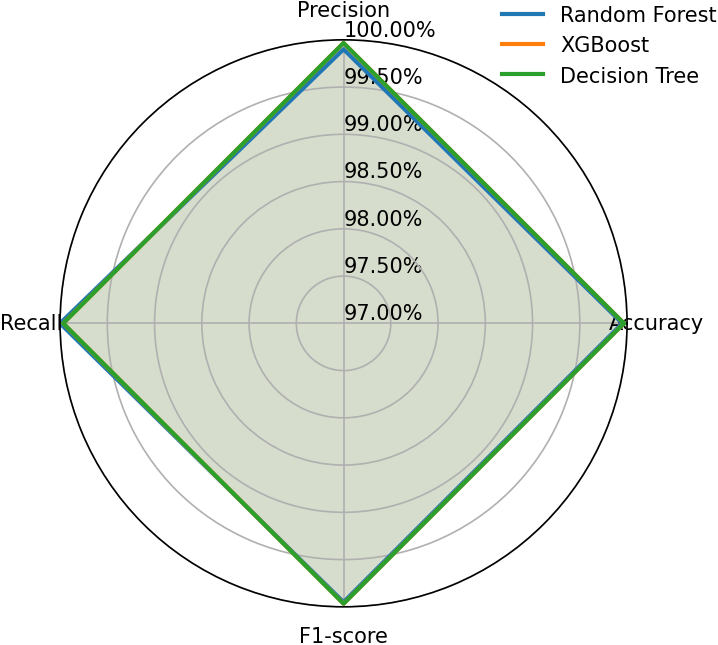}
        \caption{Excel}
        \label{????}
    \end{subfigure}
    
    \vskip\baselineskip
    \begin{subfigure}[b]{0.45\textwidth}
        \centering
        \includegraphics[width=2.5 in]{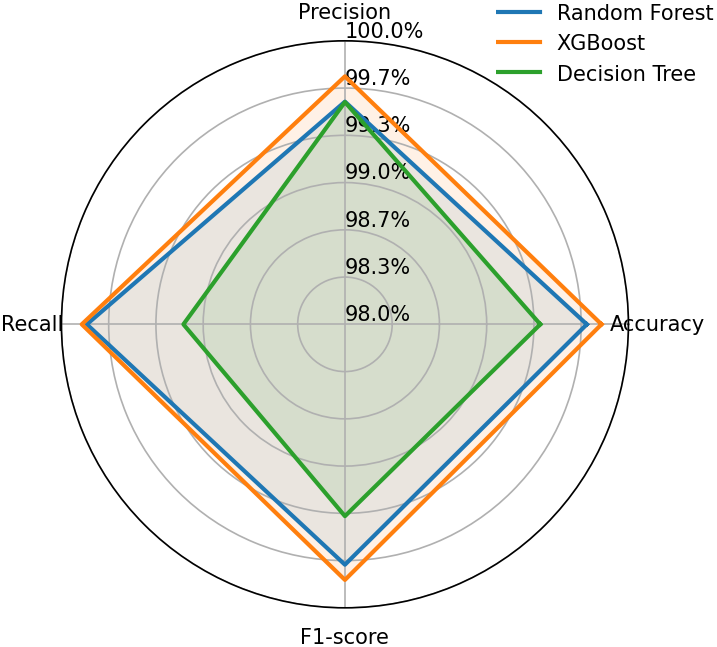}
        \caption{PDF}
        \label{????}
    \end{subfigure}
    \hfill
    \begin{subfigure}[b]{0.45\textwidth}
        \centering
        \includegraphics[width=2.5 in]{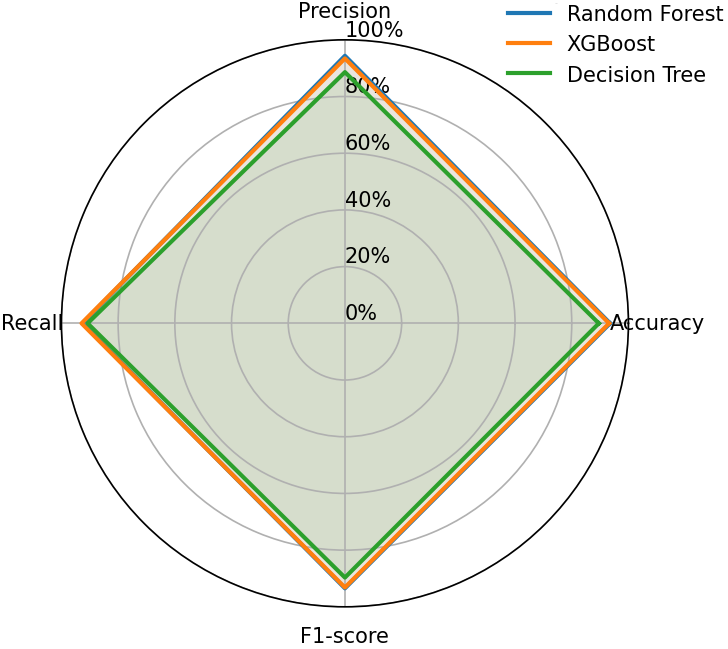}
        \caption{HTML}
        \label{????}
    \end{subfigure}

    \caption{Evaluation results of all Four Data types including Word Document, Excel, PDF and HTML}
    \label{fig:url_distribution}
\end{figure*}

\begin{figure*}[t]
  \centering
  \begin{minipage}{0.30\textwidth}
    \centering
    \includegraphics[width=\linewidth]{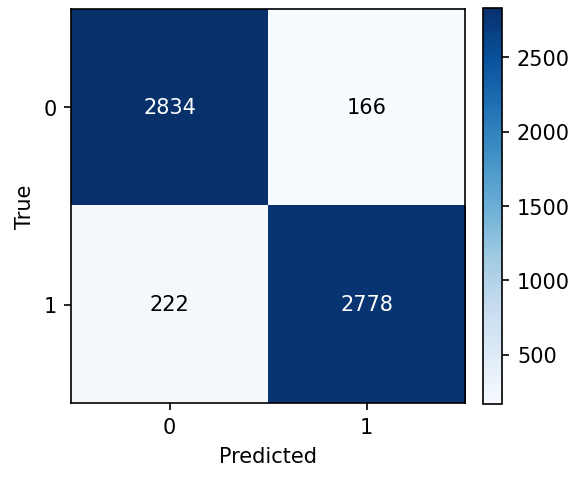}\\
    {\footnotesize (a) Random Forest}
  \end{minipage}\hfill
  \begin{minipage}{0.30\textwidth}
    \centering
    \includegraphics[width=\linewidth]{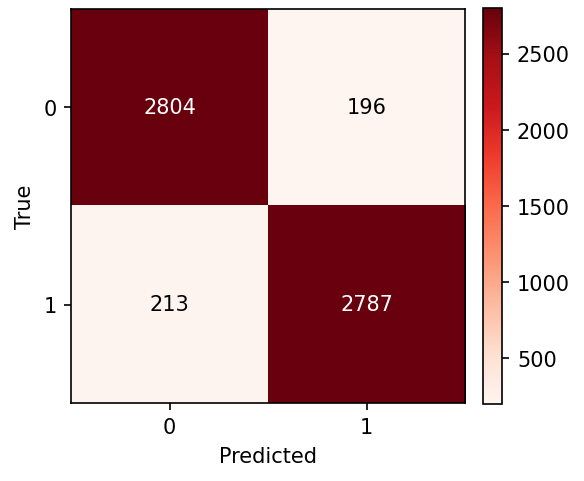}\\
    {\footnotesize (b) XGBoost}
  \end{minipage}\hfill
  \begin{minipage}{0.30\textwidth}
    \centering
    \includegraphics[width=\linewidth]{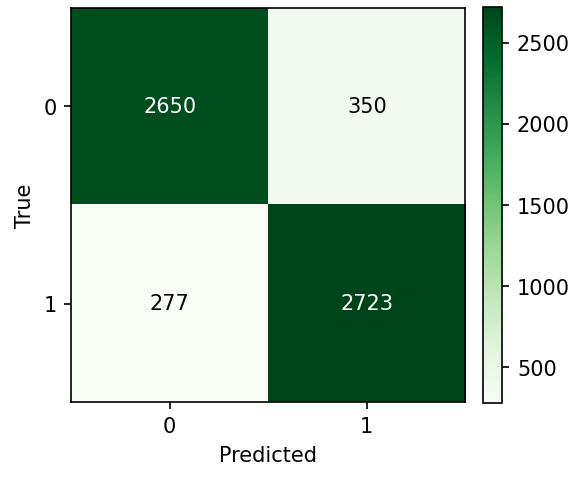}\\
    {\footnotesize (c) Decision Tree}
  \end{minipage}
  \caption{Confusion matrices of Random Forest, XGBoost, and Decision Tree classifiers for HTML files}
  \label{fig:HTML_ConfusionMatrices}
\end{figure*}

From each benign and malicious HTML sample, a set of 40 features were initially extracted, and subsequently, 13 features were selected that provided the best Accuracy, Precision, Recall and F1-score, while providing low training and testing time. By applying three classification models, XGBoost and Random Forest classifiers achieved approximately 93\% in Accuracy, Precision, and F1-score, with a Recall of 92\% which show the strength of selected features. The Decision Tree classifier achieved around 89\% across all metrics, which still indicates satisfactory classification performance. The detailed results of these models for the selected HTML features are presented in Table~\ref{Performance_Table} and visualized in Figure~\ref{fig:url_distribution} (d). The confusion matrices for all three classifiers are also illustrated in Figure~\ref{fig:HTML_ConfusionMatrices}.


 \begin{table}[H]
    \setlength{\tabcolsep}{1pt}
    \centering
    \renewcommand{\arraystretch}{1.5}
     \caption{\textcolor{black}{Per-classifier performance by data type.}}
     \label{Performance_Table}
    \footnotesize
    \begin{tabular}{>{\centering\arraybackslash}p{0.16\linewidth}
                    >{\centering\arraybackslash}p{0.18\linewidth}
                    >{\centering\arraybackslash}p{0.1\linewidth}
                    >{\centering\arraybackslash}p{0.1\linewidth}
                    >{\centering\arraybackslash}p{0.14\linewidth}
                    >{\centering\arraybackslash}p{0.14\linewidth}
                    >{\centering\arraybackslash}p{0.14\linewidth}}  
     \hline
 \rowcolor[HTML]{C0C0C0} \textbf{Data Type} & \textbf{Classifier} & \textbf{EF} & \textbf{SF} & \textbf{Precision} & \textbf{Recall} & \textbf{F1} \\
\hline
\rowcolor[HTML]{F0F0F0} \textbf{Word}  & DT & 43 & 10 & 1.0000 & 1.0000 & 1.0000  \\
               & RF & 43 & 10 & 1.0000 & 1.0000 & 1.0000  \\
               & XGBoost       & 43 & 10 & 1.0000 & 1.0000 & 1.0000 \\
\hline
\rowcolor[HTML]{F0F0F0} \textbf{Excel} & DT & 48 & 10 & 0.9995 & 0.9995 & 0.9995  \\
               & RF & 48 & 10 & 0.9995 & 0.9995 & 0.9995 \\
               & XGBoost       & 48 & 10 & 0.9998 & 0.9998 & 0.9997  \\
\hline
\rowcolor[HTML]{F0F0F0} \textbf{PDF}   & DT & 40 & 10 & 0.9930 & 0.9930 & 0.9930  \\
               & RF & 40 & 10 & 0.9958 & 0.9959 & 0.9959  \\
               & XGBoost       & 40 & 10 & 0.9963 & 0.9964 & 0.9964  \\
\hline
\rowcolor[HTML]{F0F0F0} \textbf{HTML}  & DT & 40 & 13 & 0.9113 & 0.9113 & 0.9112  \\
               & RF & 40 & 13 & 0.9386 & 0.9385 & 0.9385  \\
               & XGBoost       & 40 & 13 & 0.9379 & 0.9377 & 0.9377 \\
\bottomrule
\end{tabular}
\begin{tablenotes}
          \item \footnotesize \textbf{Note}: DT= Decision Tree, RF= Random Forest. Metrics are macro-averaged.    
    \end{tablenotes}
\end{table}


\subsubsection{QR code Evaluation}

\begin{figure}[t]
  \centering
  \begin{minipage}{0.4\linewidth}   
    \centering
    \includegraphics[width=1\linewidth]{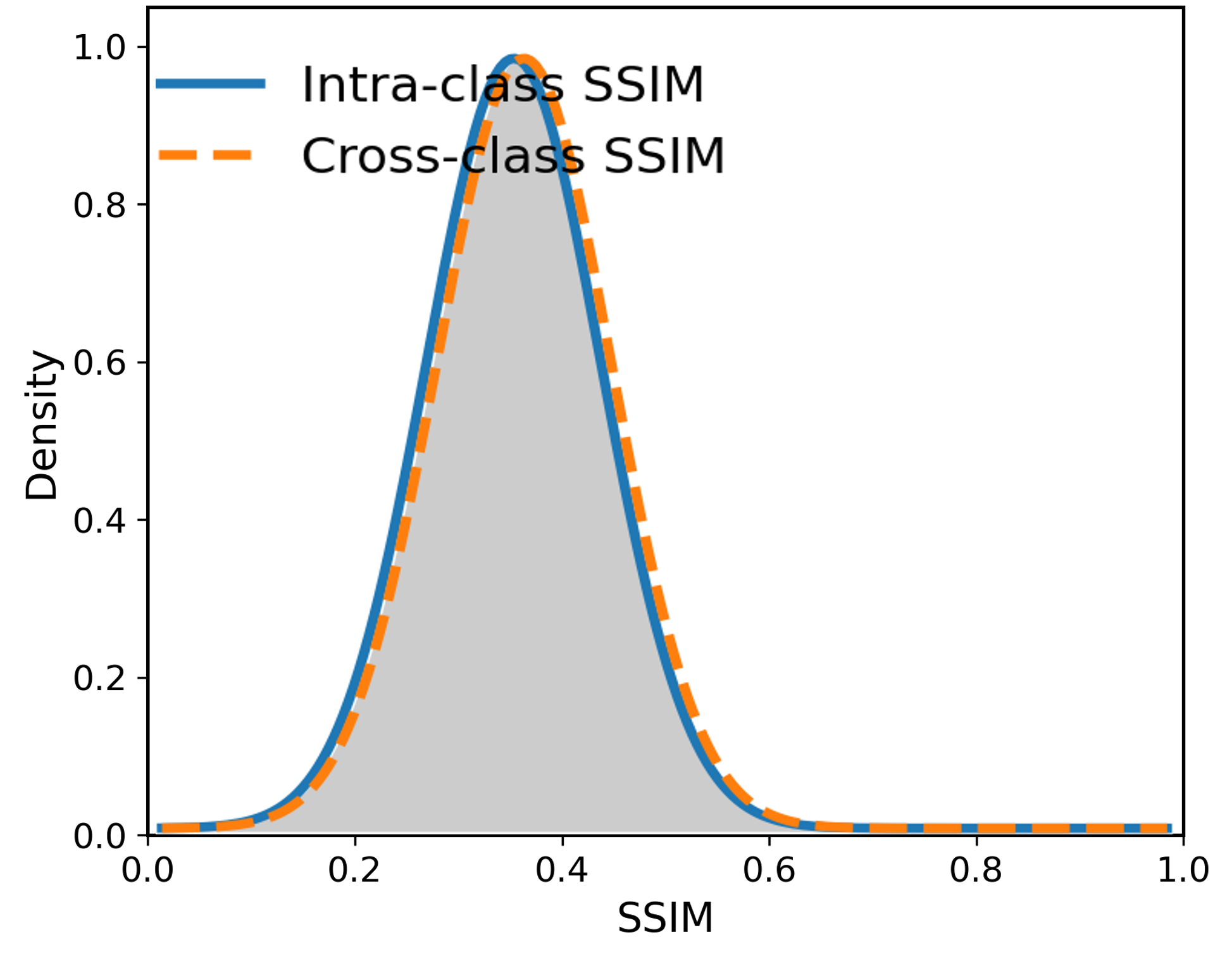}\\ 
    {\footnotesize (a) Structural Similarity (SSIM) }
  \end{minipage}
  \hspace{0.01\linewidth}            
  \begin{minipage}{0.4\linewidth}
    \centering
    \includegraphics[width=1\linewidth]{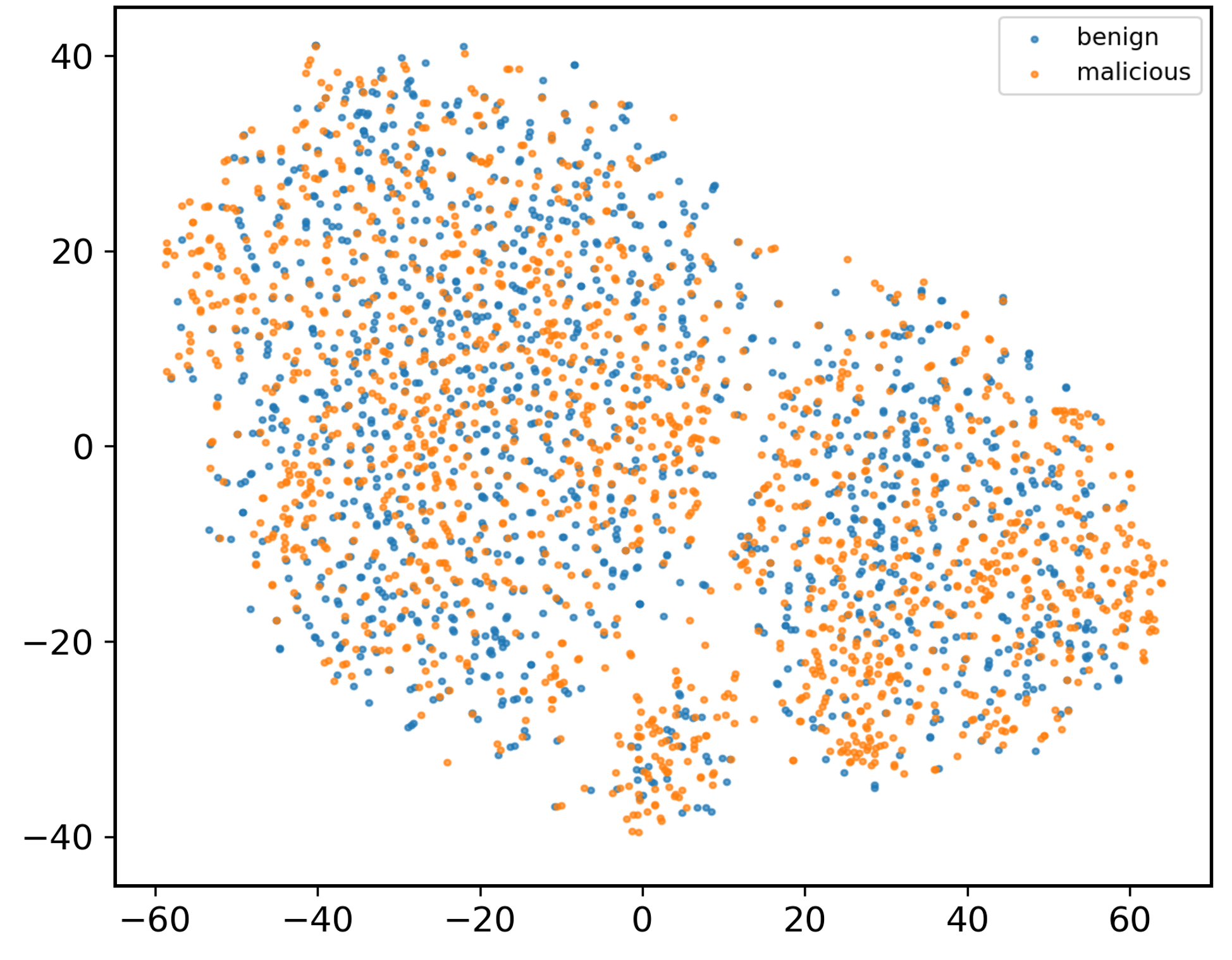}\\
    {\footnotesize (b) t-SNE visualization}
  \end{minipage}
  \caption{Structural and perceptual similarity analysis of benign and malicious QR codes.}
  \label{StructuralCNN_Analysis}
\end{figure}

In order to evaluate the QR codes samples that were already generated from the malicious and benign URLs, we firstly followed an image processing approach by applying a basic CNN model to the QR code images. The corresponding performance results are provided in Table \ref{Comparative_results_withLLMnew}. As shown in this table, the CNN model accuracy is moderate, despite being trained on a large dataset, which proves that the image-based approach has some limitations in detection phase. To further investigate these limitations, additional visual and structural similarity analyses were performed. The results of these analyses are presented in Figure \ref{StructuralCNN_Analysis}.

The separability analysis shows that benign and malicious QR codes have highly similar visual characteristics, making them almost indistinguishable, which leads to noticeable overlap between the two classes. The quantitative similarity metrics which include a silhouette score of 0.002 and center-to-intra-class ratio of 0.067, also proves this overlap. Moreover, the value of structural similarity index (SSIM) between benign and malicious QR images is 0.34-0.35 which indicates that their pixel-level spatial arrangements contain identical visual textures. Therefore, these analyses indicates that image-based methods face challenges because QR encoding creates identical visual patterns for both malicious and benign URLs through its black-and-white module structure.

The next step that we followed involved analyzing the URL strings corresponding to the QR codes to evaluate the effectiveness of benign and malicious URL samples in the detection phase from a lexical perspective. In order to examine different models, several lightweight language models (LLMs), including BERT-Tiny, DeBERTa-v3, ModernBERT, and DeepSeek-R1, were applied to the URL dataset in this experiment. The results presented in Table \ref{Comparative_results_withLLMnew} proves that LLM-based methods (URL lexical analyses) achieve better performance compared to the CNN model, as reflected by their higher Precision, Recall, and F1-score values. Moreover, the training time of these models is significantly lower than image based approach, since they process character level representations and tokenized text rather than images.

\begin{figure}[t]
  \centering
  \includegraphics[width=0.6\linewidth]{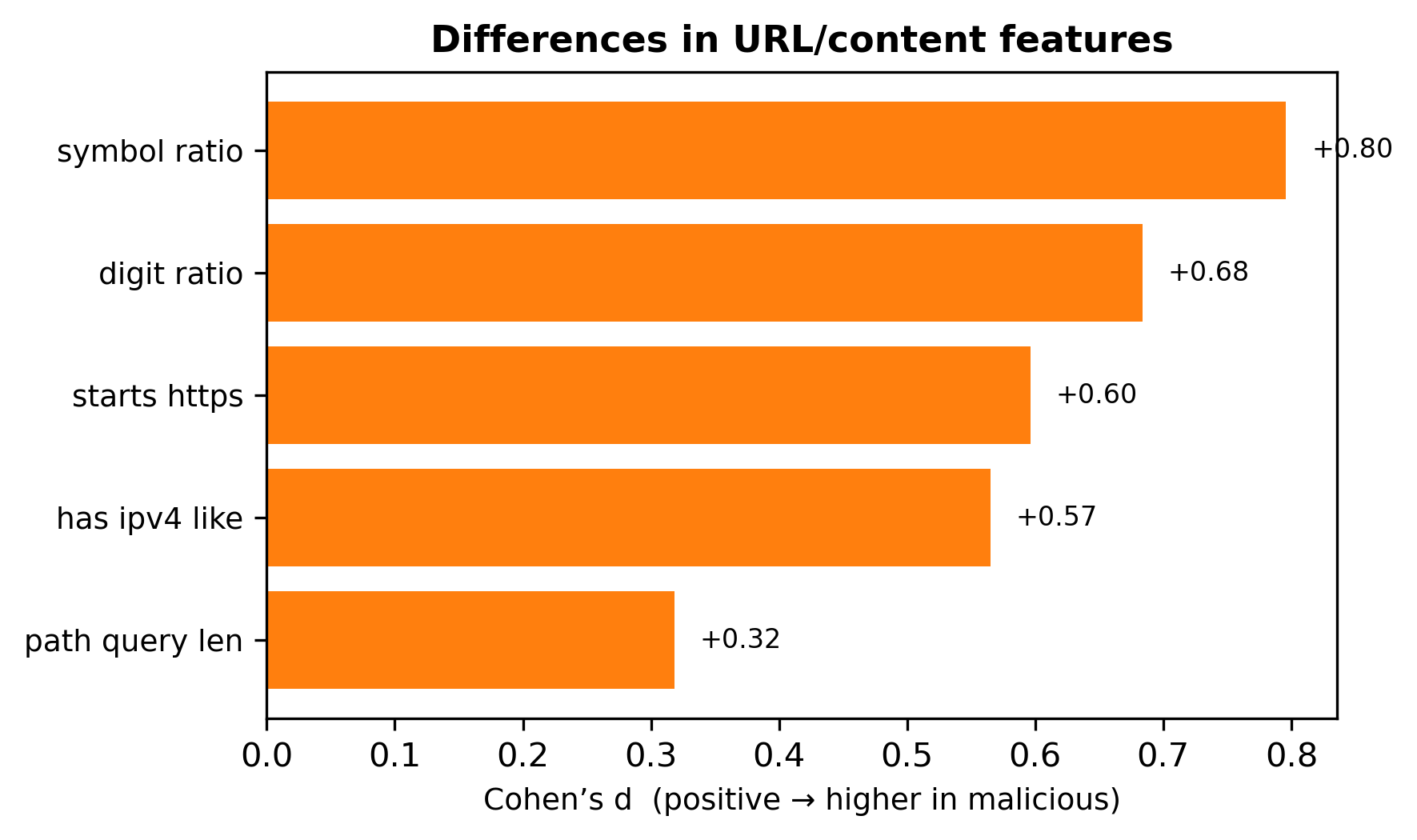}
  \caption{Lexical/structural differences between benign and malicious QR-embedded URLs.}
  \label{fig:url_features_effect_sizes}
\end{figure}

The analysis investigated whether the discriminative data exists within text encoding or image content by analyzing URL statistics from decoded QR codes. Figure \ref{fig:url_features_effect_sizes} demonstrates that benign and malicious URLs produce distinct behavioral and lexical patterns. The malicious samples show significant increases in digit ratios at +0.68 Cohen’s d and symbol ratios at +0.80 and IPv4-like pattern ratios at +0.57 and longer query paths and higher HTTPS start probability at +0.60. The visual appearance of QR codes remains similar yet their embedded URL strings contain recognizable textual and structural characteristics which text-based and language model approaches can successfully identify.

\begin{table}[!h]
    \setlength{\tabcolsep}{1pt}
    \centering
    \scriptsize
    \caption{Performance results for malicious QR code detection using image-based approaches and recent lightweight LLMs applied to URL-based analysis.}
    \renewcommand{\arraystretch}{1.5}
    \begin{tabular}{
        >{\centering\arraybackslash}p{0.23\linewidth}
        >{\centering\arraybackslash}p{0.1\linewidth}
        >{\centering\arraybackslash}p{0.1\linewidth}
        >{\centering\arraybackslash}p{0.12\linewidth}
        >{\centering\arraybackslash}p{0.11\linewidth}
        >{\centering\arraybackslash}p{0.1\linewidth}
        >{\centering\arraybackslash}p{0.1\linewidth}
        >{\centering\arraybackslash}p{0.1\linewidth}
    }
        \hline
        \rowcolor[HTML]{C0C0C0}
        Detection Classifier & Approach & Train Time (s) & Inference Time (s) & Dataset Size & Precision (M) & Recall (M) & $\boldsymbol{F_1}$ Score (M) \\
       \hline
    CNN &  Image-Based & 61,485 & 284  &  1,000,000 & 0.9301 & 0.8401  &  0.8828 \\ \hline   
  
 \rowcolor[HTML]{F0F0F0}  BERT-Tiny \cite{turc2019well}  & URL-Based & 1,693 &  28
 &   1,072,659 & 0.9856 &  0.9860 &  0.9858 \\  

 \rowcolor[HTML]{F0F0F0}  DeBERTa-v3 \cite{he2021debertav3}  & URL-Based & 9,196 &  129
 &  1,072,659  & 0.9917 &  0.9924 &  0.9920 \\  

 \rowcolor[HTML]{F0F0F0}  ModernBERT \cite{warner2024smarter}  & URL-Based & 8,633  &  111
 &   1,072,659 & 0.9939 &  0.9922 &  0.9930 \\  
 
  \rowcolor[HTML]{F0F0F0}  DeepSeek-R1 (Distill-Qwen) \cite{guo2025deepseek}  & URL-Based &  12,861 &  252  &  1,072,659  & 0.9606 &  0.9611 &  0.9609 \\  \hline
   
    \end{tabular}         \label{Comparative_results_withLLMnew}
    \begin{tablenotes}
          \item \footnotesize \textbf{Note:}  \textbf{M}: macro average across classes. 
    \end{tablenotes}
\end{table}

\section{Conclusion}
\label{Nazgol_Malware_conclusion}
The security systems face ongoing difficulties because phishing and malware distribution continue to use email attachments and QR codes as their primary attack methods. The development of unified datasets which include diverse and representative content requires immediate attention because it will help improve malicious content detection systems for various file formats. Our research established a comprehensive malicious-attachment dataset which includes five common file types (Microsoft Word documents, Excel spreadsheets, PDF files, HTML pages and QR codes) with both malicious and benign samples. 

In order to analyze the first four data types, this paper proposed feature extraction and feature selection pipelines, which lead to have a set of features for each data type. The final set of features were tested by applying Decision Tree, Random Forest and XGBoost classification models. The detection accuracy remained high for all document types according to the results which proved that the extracted and selected features worked well. In order to analyze the QR code samples was done by following two different approaches including a CNN model for image-based analysis and also using lightweight Large Language Models (LLMs) for URL-based evaluation. The results show that the LLMs outperformed the CNN in all evaluation metrics because they used text-based semantic analysis to understand URLs better.

The research introduces the first complete dataset which unites static features with machine learning detection methods for malicious attachments and QR code phishing attacks. The dataset offers researchers and practitioners an essential tool to create explainable detection systems which operate efficiently and at scale for protecting against real-world phishing attacks. The complete dataset will become available through the Canadian Institute for Cybersecurity (CIC) Dataset Portal to support researchers who want to improve their skills in malicious document and QR-code analysis.

\section*{Data Availability}
The dataset generated in this study is publicly released through the Canadian Institute for Cybersecurity’s official repository at:  

https://www.unb.ca/cic/datasets/trap4phish2025.html

\section*{Acknowledgements}
The authors express their gratitude to the anonymous reviewers for their valuable feedback. Additionally, the authors sincerely appreciate the support received from the Canadian Institute for Cybersecurity (CIC).

\bibliographystyle{IEEEtran}
\bibliography{IEEEabrv,PAN-DNN_Nazgol}

\end{document}